%% file: main.tex
\documentclass[prd,aps,showpacs,tightenlines, 10pt, reprint,nofootinbib,superscriptaddress,floatfix,twocolumn]{revtex4-2}
\input{packages.tex}

\input{commands.tex}

\begin{document}

%%%%%%%%%%%%%%%%%%%%%%%% Title page %%%%%
\title{Circularization vs. Eccentrification in Intermediate Mass Ratio Inspirals inside Dark Matter Spikes}
\author{Niklas Becker}
\email{nbecker@itp.uni-frankfurt.de} 
\affiliation{Institute for Theoretical Physics, Goethe University, 60438 Frankfurt am Main, Germany}

\author{Laura Sagunski}
\email{sagunski@itp.uni-frankfurt.de} 
\affiliation{Institute for Theoretical Physics, Goethe University, 60438 Frankfurt am Main, Germany}

\author{Lukas Prinz}
\email{lprinz@itp.uni-frankfurt.de} 
\affiliation{Institute for Theoretical Physics, Goethe University, 60438 Frankfurt am Main, Germany}

\author{Saeed Rastgoo}
\email{srastgoo@yorku.ca} 
\affiliation{Department of Physics and Astronomy, York University, Toronto, Ontario, M3J 1P3, Canada}
%\author{}
%\email{}
%\affiliation{\frankfurt}
%\author{}
%\email{}
%\affiliation{\york}
\date{\today}

\begin{abstract}
Inspirals of an Intermediate Mass Black Hole (IMBH) and a solar mass type object will be observable by space based gravitational wave detectors such as The Laser Interferometer Space Antenna (LISA). A dark matter overdensity around an IMBH -- a dark matter spike -- can affect the orbital evolution of the system.  We consider here such Intermediate Mass Ratio Inspirals on eccentric orbits, experiencing dynamical friction of the dark matter spike. We find that by including the relative velocities of the dark matter particles, the dynamical friction tends to circularize the orbit, in contrast to previous inquiries. We derive a general condition for circularization or eccentrification for any given dissipative force. In addition to the dephasing, we suggest using the circularization rate as another probe of the dark matter spike. Observing these effects would be an indicator for the particle nature of dark matter.
\end{abstract}
%\preprint{}
%\pacs{gravitational waves, dark matter, \SRC{pacs are usually alphanumerical codes}}
\maketitle

%%%%%%%%%%%%%%%%%%%%%%%%%%%%%%%%%%%%%%%%
\section{Introduction \label{sec:intro}} 
\input{introduction.tex}

%%%%%%%%%%%%%%%%%%%%%%%%%%%%%%%%%%%%%%%%

%%%%%%%%%%%%%%%%%%%%%%%%%%%%%%%%%%%%%%%%
\section{IMRI Modeling \label{sec:equations}}

\input{equations.tex}

%%%%%%%%%%%%%%%%%%%%%%%%%%%%%%%%%%%%%%%%
%%%%%%%%%%%%%%%%%%%%%%%%%%%%%%%%%%%%%%%%
\section{Results \label{sec:results}}
\input{results.tex}

\section{Analysis \label{sec:analysis}}
\input{analysis.tex}
%%%%%%%%%%%%%%%%%%%%%%%%%%%%%%%%%%%%%%%%

\section{Conclusions \label{sec:concl}} 
\input{conclusions.tex}

%%%%%%%%%%%%%%%%%%%%%%%%%%%%%%%%%%%%%%%%

%%%%%%%%%%%%%%%%%%%%%%%%%%%%%%%%%%%%%%%%

\begin{acknowledgments}
We thank Bradley Kavanagh and the two referees for helpful discussions and comments on the initial manuscript. N.B. and L.S. acknowledge support by the Deutsche Forschungsgemeinschaft (DFG, German Research Foundation) through the CRC-TR 211 'Strong-interaction matter under extreme conditions'– project number 315477589 – TRR 211. S. R. acknowledges the support of the Natural Sciences and
Engineering Research Council of Canada (NSERC), funding
reference numbers RGPIN-2021-03644 and DGECR-2021-00302 
\end{acknowledgments}
%%%%%%%%%%%%%%%%%%%%%%%%%%%%%%%%%%%%%%%%

%%%%%%%%%%%%%%%%%%%%%%%%%%%%%%%%%%%%%%%%
\appendix
\input{appendix.tex}
%%%%%%%%%%%%%%%%%%%%%%%%%%%%%%%%%%%%%%%%

%%%%%%%%%%%%%%%%%%%%%%%% Bibliography %%%%%
\bibliographystyle{apsrev4-1}
\bibliography{biblio}{}
%%%%%%%%%%%%%%%%%%%%%%%%%%%%%%%%%%%%%%%%
\end{document}

%% file: packages.tex
\usepackage[utf8]{inputenc} 

\usepackage{acronym}
\usepackage{amsmath}
\usepackage{amsthm}  % theorems etc
\usepackage{amssymb}
\usepackage{mathtools}
\usepackage{phaistos}
\usepackage{graphicx}
\usepackage{subcaption}
\usepackage[section]{placeins}

\usepackage{physics} % /dv /pdv
\usepackage{IEEEtrantools}  % equations
\usepackage{xspace}
\usepackage{cancel}
\usepackage[normalem]{ulem} % strikeout 
\usepackage{epsfig}
\usepackage{multirow} 
\usepackage{color}
\usepackage[dvipsnames]{xcolor}
\usepackage{rotating}
\usepackage[export]{adjustbox} %To adjust subplots at the top.
\usepackage{floatrow}
\newfloatcommand{capbtabbox}{table}[][\FBwidth]
\usepackage{tabularx}
\usepackage[pdfencoding=auto, psdextra]{hyperref}
\usepackage{units}
\usepackage[T1]{fontenc}
\usepackage[normalem]{ulem} %For crossing out words.
\usepackage{aas_macros}

%% file: commands.tex
\newcommand{\nn}{\nonumber}

\newcommand{\Msun}{M_\odot}

\newcommand{\mchirp}{\mathcal{M}_c}
\newcommand{\pc}{\text{pc}}

\newcommand{\dm}{\text{dm}}
\newcommand{\isco}{\text{isco}}
\renewcommand{\sp}{\text{spike}}
\newcommand{\E}{\mathcal{E}}
\newcommand{\F}{\mathcal{F}}
\renewcommand{\v}[1] {\mathbf{#1}}
\DeclarePairedDelimiterX{\avg}[1]{\langle}{\rangle}{#1}
\makeatletter
\let\oldavg\avg
\def\avg{\@ifstar{\oldavg}{\oldavg*}}  

\renewcommand{\eqref}[1]{Eq.~(\ref{#1})}
\newcommand{\neqref}[1]{(\ref{#1})}
\newcommand{\figref}[1]{Fig.~\ref{#1}}
\newcommand{\lcdm}{$\Lambda$CDM}

%% file: introduction.tex
%%%%%%%%%%%%%%%%%%%%%%%%%%%%%%%%%%%%%%%%
%\subsection{Context} 
%%%%%%%%%%%%%%%%%%%%%%%%%%%%%%%%%%%%%%%%
The first detection of gravitational waves (GWs) has opened a fundamentally new window into the Universe. The Laser Interferometer Gravitational-Wave Observatory (LIGO) collaboration has seen the first binary black hole merger, and, together with the Virgo collaboration, has already collected a sizable catalogue of binary black hole and neutron star mergers by now \cite{LIGOScientific:2016aoc, LIGOScientific:2021djp}. These compact binary mergers allow new and unprecedented tests of General Relativity and matter at extremely high densities \cite{LIGOScientific:2020tif, LIGOScientific:2018cki}. In addition to ground-based detectors such as LIGO and Virgo, there are several space-based observatories planned, such as LISA \cite{LISA:2017pwj}, Taiji \cite{10.1093/nsr/nwx116} and TianQuin~\cite{TianQin:2015yph}. 

Meanwhile, the nature of dark matter continues to elude direct and indirect detection probes \cite{Bertone:2004pz, XENON:2018voc}. First proposed to explain galactic rotation curves, and integral to the success of the standard cosmological \lcdm{} model, the hunt for dark matter has been going on for decades, with no fruitful results. The \lcdm{} model utilizes cold, collisionless dark matter particles virializing into halos and seeding the formation of structures in the universe. On small scales, the effects of dark matter are more uncertain and a plethora of models have been proposed \cite{Bertone:2018krk}. 

While LIGO mostly observes solar mass binary mergers, LISA will be able to observe IMBHs with masses ranging from $10^2 \sim 10^6\Msun$. IMBHs have been detected, but their origin and evolution is not well understood as of now \cite{Mezcua:2017npy}. Around these IMBHs, on very small scales, a dark matter halo could grow adiabatically into a dark matter \textit{spike} \cite{Gondolo:1999ef, Sadeghian:2013laa}. These spikes have an extremely high local density and would gravitationally interact with any object passing by. During an Intermediate Mass Ratio Inspiral (IMRI), where a stellar mass object inspirals onto an IMBH, the dark matter spike can leave its imprint by modifying the orbital evolution.

This has first been explored in \cite{Eda:2013gg, Eda:2014kra}, where the authors predicted a dephasing of the GW signal due to dynamical friction which the secondary object experiences while passing through the dark matter spike \cite{Chandrasekhar:1943ys}. This slows down the object and results in a faster inspiral, which would be observable in the phase evolution of the GW signals that can be detected by LISA \cite{Barausse:2014tra, Coogan:2021uqv}. 

Additionally, if the secondary object is also a black hole, it will accrete (i.e., absorb) some of the dark matter as it passes through the spike. This was first explored in \cite{Macedo:2013qea} and later \cite{Yue:2017iwc}, where the accretion effects were found to be subdominant to dynamical friction effects, but still important on the long timescales involved. Then, \cite{Yue:2019ozq, Cardoso:2020iji} looked at eccentric orbits, instead of using the circular approximation that was employed before, and found there to be an eccentrification of the orbits. This would mean that the circular approximation cannot be utilized, and that we should expect most IMRIs in dark matter spikes to be highly elliptical. 

Meanwhile, \cite{Kavanagh:2020cfn} developed a model that promoted the dark matter spike from a background actor to an integral part of the evolution with the \textit{halo feedback} model: As the secondary object passes through the spike, it loses momentum, which is transferred into the dark matter halo and locally depletes it. This results in a lower dark matter density and less dynamical friction effects, and thus in a longer inspiral compared to the static halo case. As the object inspirals, the depleted region moves inward with it and the outer region is refilled, leaving the spike itself mostly intact. Nevertheless, this halo feedback model relies on the circular orbit approximation, which according to \cite{Yue:2019ozq, Cardoso:2020iji} would be an unrealistic scenario. \\
%%%%%%%%%%%%%%%%%%%%%%%%%%%%%%%%%%%%%%%%
%\subsection{Motivation}
%%%%%%%%%%%%%%%%%%%%%%%%%%%%%%%%%%%%%%%%

The motivation in this paper is to model IMRIs on elliptical Keplerian orbits with GW emission, dynamical friction, and include the relative velocities by means of the phase space description of the dark matter halo in one consistent framework. We find orbital circularization instead of eccentrification through dynamical friction.  We explore the evolution and GW signal from different initial conditions and different model parameters. We derive a condition for eccentrification and circularization for general dissipative forces acting on the secondary object. We derive the circularization rate depending on the dark matter spike properties, which can be used as another probe of the spike and thus dark matter particle properties.

The structure of the paper is as follows. In section \ref{sec:equations}, we explain the theoretical framework to model the orbital evolution of the IMRI and its GW emission. In section \ref{sec:results} we present our numerical results. We analyse them in section \ref{sec:analysis}. Finally, we draw our conclusions in section \ref{sec:concl}. 
%\LS{Mention the appendices.} 

Throughout the paper we adopt geometrized units with $c = G = 1$.

%% file: equations.tex
\subsection{Dark Matter Spike}
We consider an IMRI in which the central mass $m_1$ is assumed to be surrounded by a static, spherically symmetric dark matter spike. This spike can develop by adiabatic growth of the central black hole. Initially, a black hole seed grows by accretion of the surrounding halo, and the slow increase of the potential concentrates the dark matter particles into a density spike \cite{Gondolo:1999ef, Ullio:2001fb, Sadeghian:2013laa}. 

The existence of spikes around black holes is not certain. They might be disrupted by processes such as major mergers, and the models require the black holes to be in the center of the dark matter halo. The dark matter particles have to be non-annihilating and rather cold \cite{Coogan:2021uqv}. Nevertheless, their existence would be an indicator of the particle nature of dark matter and could reveal much about their host black hole's history \cite{Ullio:2001fb}. 

We adopt the description proposed in \cite{Coogan:2021uqv} and describe the dark matter density around the IMBH by a simple power law
\begin{equation}
\label{eq:rho_dm}
    \rho_\dm(r) = \rho_6 \left(\frac{r_6}{r}\right)^{\alpha_\sp}, \quad  r_\text{in} < r < r_\sp 
\end{equation}
with the radius from the central black hole $r$ and the reference radius $r_6 = 10^{-6}$pc. Following \cite{Sadeghian:2013laa}, the inner radius is chosen to be $r_\text{in}=4m_1$. The spike radius $r_\sp$ is the maximal radius of the spike, which can be obtained by comparing the gravitational influence of the central black hole to the total spike mass \cite{Eda:2014kra}. The range of the power law index is $1 < \alpha_\sp < 3$. Different origins of the spike can give different values for $\alpha_\sp$, such as $\alpha_\sp=7/3$ for an NFW halo forming a spike \cite{Eda:2014kra}, $\alpha_\sp=7/4$ for self-interacting dark matter (SIDM) forming a spike \cite{Shapiro:2014oha}, or $\alpha_\sp=9/4$ for a dark matter spike around Primordial Black Holes \cite{Boudaud:2021irr}. 

The description found in other literature with $\rho_\dm(r) = \rho_\sp (\frac{r_\sp}{r})^{\alpha_\sp}$ can be recovered using \cite{Coogan:2021uqv} 
\begin{align*}
    \rho_\sp &=  \left( \rho_6 r_6^{\alpha_\sp} (km_1)^{-\alpha_\sp/3}  \right)^{3/(3-\alpha_\sp)}\\
    r_\sp &= \left( \frac{km_1}{\rho_\sp} \right)^{1/3} \\
    k &= \frac{3-\alpha_\sp}{2\pi} 0.2^{3-\alpha_\sp}. \nn  
\end{align*}
The dark matter particles in the halo can be described by an equilibrium phase space distribution function $f=dN/d^3rd^3v$, describing the number density per phase space volume. In our case, since the halo is spherically symmetric,  $f = f(\E)$, where $\E$ is the relative energy per unit mass
\begin{equation}
    \E(r,v) = \Psi(r) - \frac{1}{2}v^2
\end{equation}
with the relative Newtonian gravitational potential $\Psi(r)$. Close to the black hole, this is simply $\Psi(r) = \frac{m_1}{r}$. Gravitationally bound particles are those with $\E > 0$. 

For a given spherically symmetric density profile $\rho(r)$, the distribution function $f(\E)$ can be obtained by the Eddington inversion procedure \cite{1987gady.book}. For the power law spike, this is given by
\begin{align}
\label{eq:f_powerlaw}
    f_\sp(\E) =& \frac{\alpha_\sp(\alpha_\sp-1)}{(2 \pi)^{3/2}} \rho_6 \left(\frac{r_6}{m_1}\right)^{\alpha_\sp} \nn \\
     & \times \frac{\Gamma(\alpha_\sp - 1)}{\Gamma(\alpha_\sp - \frac{1}{2})} \E^{\alpha_\sp-3/2}
\end{align}
with the Gamma function $\Gamma$. This gives us a lower bound on $\alpha_\sp > 1$. The upper bound $\alpha_\sp < 3$ is derived from the requirement that the enclosed mass is finite.

The density for a given distribution function is recovered through
\begin{equation}
\label{eq:rho_f}
    \rho(r) = 4\pi \int_0^{v_{\text{max}}(r)} v^2 f\left(\Psi(r) - \frac{1}{2}v^2 \right) dv
\end{equation}
with the escape velocity at radius $r$ given by $v_{\text{max}}(r)=\sqrt{2\Psi(r)}$.

\subsection{Orbital Evolution}
\begin{figure}
    \centering
    \includegraphics[width=0.9\textwidth]{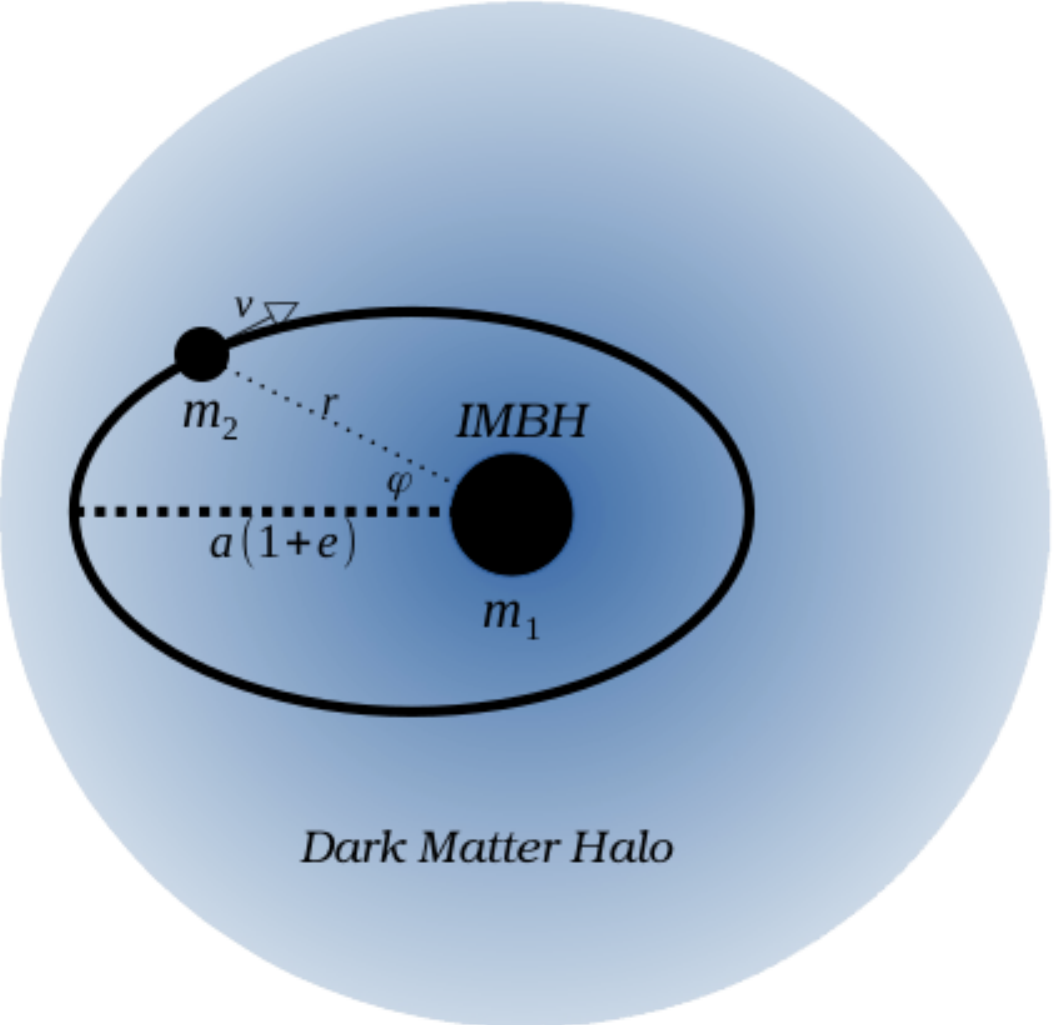}
    \caption{A sketch of the Keplerian system with masses $m_1 \gg m_2$, semimajor axis $a$, and eccentricity $e$ inside the dark matter halo $\rho_\dm$. In \eqref{eq:Kepler_r}, we have $\phi = \pi - \varphi$, such that $\phi = 0$ is the periapsis, the closest point in the orbit, and $\phi = \pi$ the apoapsis, the farthest point in the orbit. }
    \label{fig:sketch}
\end{figure}

The IMRI system consists of a central mass $m_1$ and a secondary object $m_2$, both of which are assumed to be Schwarzschild black holes for simplicity as depicted in \figref{fig:sketch}. The secondary object is assumed to be on a Keplerian orbit around the central mass. The system emits GWs that might be observable by future GW detectors such as LISA. Through this GW emission and other dissipative forces, the secondary object loses orbital energy and angular momentum, leading to an inspiral orbit.

\subsubsection{Keplerian Orbit}
The secondary object is assumed to be on a Keplerian orbit around the central mass. Here, we ignore the additional dark matter contribution to the total and reduced mass $\mu$ of the Keplerian system and assume $m=m_1 + m_2$, $\mu = \frac{m_1 m_2}{m}$, respectively. This is a decent approximation, since we are looking at systems close to inspiral.  These have small orbital separations that are gravitationally dominated by the central black hole, because the total enclosed mass of the spike up to the location of the orbiting object is much smaller than the mass of the central black hole, $m_\dm(r=10^5r_{\isco}) \ll m_1$. Here, $r_\isco$ refers to the radius of the innermost stable circular orbit for massive objects, which is $r_\isco=6m_1$ for a Schwarzschild black hole. Following \cite{Dai:2021olt}, the inclusion of the gravitational influence of the spike as a perturbative force would primarily lead to orbital precession, which we neglect in this paper. 

Any Keplerian orbit can be described by two parameters, the semimajor axis $a$ and the eccentricity $e$. For a bound orbit, we have $0 \leq e < 1$, where $e=0$ describes a circular orbit.

The orbital energy is given by \cite{Maggiore:2007ulw}
\begin{equation}
    E_{\text{orb}} = -\frac{m\mu}{2a}
\label{eq:E_orbit}
\end{equation}
and the angular momentum $L_{\text{orb}}$ is given by the following relation
\begin{equation}
    e^2 - 1 = \frac{2E_{\text{orb}}L_{\text{orb}}^2}{m^2 \mu^3}. 
\label{eq:L_orbit}
\end{equation}
Throughout one orbit, the radius and the velocity of the orbiting object at the true anomaly $\phi$ can be obtained by the relations
\begin{align}
    r =& \frac{a(1-e^2)}{1+ e \cos{\phi}} \label{eq:Kepler_r}\\
    v^2 =& m \left( \frac{2}{r} - \frac{1}{a} \right), \label{eq:Kepler_v}
\end{align}
while the mean orbital frequency 
%\LP{How is it defined? Also, $F$ is used later for forces, maybe use a different symbol here for the mean frequency?}\NBc{Both are unfortunately pretty standard. It should be clear from context and for the force we always have a subscript or parenthesis $F(r,v)$}\LS{To make this clear, maybe we can mention this explicitly when we introduce the force for the first time?} 
%
is given analogously to the circular case by
\begin{equation}
\label{eq:Kepler_F}
        \mathcal{F} = \frac{1}{2\pi} \sqrt{\frac{m}{a^3}}.
\end{equation}

\subsubsection{Dissipative Forces}
The orbit is assumed to lose energy on a secular timescale much larger than the orbital timescale. This assumption allows us to use the Keplerian orbits to calculate the forces acting on the object. Over many orbits, these forces then lead to a change in the orbital parameters. 
To model the dissipative forces, we use the force term $F(r, v)$ depending on the distance $r$ and the velocity $v$ of the secondary object.

The energy and angular momentum loss for a given dissipative force are obtained by averaging over one orbit with orbital period $T$ \cite{Yue:2019ozq},
\begin{align}
    \avg{\dv{E}{t}} =& \int_0^T \frac{\mathrm{d}t}{T} \dv{E}{t} =- \int_0^T \frac{\mathrm{d}t}{T} F(r,v)v,  \label{eq:avgdEdT} \\
    \avg{\dv{L}{t}} =& \int_0^T \frac{\mathrm{d}t}{T} \dv{L}{t} = - \sqrt{ma(1-e^2)}\int_0^T \frac{\mathrm{d}t}{T} \frac{F(r,v)}{v}. \label{eq:avgdLdT}
\end{align}
The second equation is derived in appendix \ref{sec:app_AML}. These integrals can be computed by the following relation, which is valid for an arbitrary function $G(r,v)$ \cite{Maggiore:2007ulw},
\begin{equation}
    \int_0^T \frac{\mathrm{d}t}{T} G(r(t),v(t)) = (1-e^2)^{\frac{3}{2}}\int_0^{2\pi} \frac{\mathrm{d}\phi}{2\pi}\frac{G(r(\phi), v(\phi))}{(1 + e \cos{\phi})^{2}}  
\end{equation}
and with the help of Eqs.~\neqref{eq:Kepler_r} and \neqref{eq:Kepler_v}.

Thus, for a given force $F(r,v)$, we can compute the energy and angular momentum loss either analytically or numerically using Eqs. \neqref{eq:avgdEdT} and \neqref{eq:avgdLdT}. 

The specific effects considered here are GW emission loss and dynamical friction. Each can be modeled as a force and lead to a loss of orbital energy over secular timescales
\begin{equation}
    \label{eq:dEorb_dt}
    \dv{E_\text{orb}}{t} = \avg{\dv{E_\text{gw}}{t}} + \avg{\dv{E_\text{df}}{t}} .
\end{equation}
Similarly, the angular momentum of the orbit dissipates over secular timescales as 
\begin{equation}
\label{eq:dLorb_dt}
    \dv{L_{\text{orb}}}{t} = \avg{\dv{L_{\text{gw}}}{t}} + \avg{\dv{L_{\text{df}}}{t}} .
\end{equation}
\\

\noindent\textbf{Gravitational Waves}

The GW emission terms are given by \cite{Maggiore:2007ulw}
\begin{align}
    \avg{\dv{E_{\text{gw}}}{t}} =& - \frac{32}{5} \frac{\mu^2 m^3 }{a^5} \frac{1 + \frac{73}{24}e^2 +\frac{37}{96}e^4}{(1-e^2)^{7/2}}, \label{eq:dE_gw}\\
    \avg{\dv{L_{\text{gw}}}{t}} =& - \frac{32}{5} \frac{\mu^2 m^{5/2} } {a^{7/2}} \frac{1 + \frac{7}{8}e^2}{(1-e^2)^2} .\label{eq:dLgw}
\end{align}
\\

\noindent\textbf{Dynamical Friction}

The dynamical friction is given by the Chandrasekhar equation \cite{Chandrasekhar:1943ys, Kavanagh:2020cfn}
\begin{equation}
\label{eq:F_df}
    F_{\text{DF}}(r, v) = \frac{4\pi m_2^2 \rho_\dm(r) \xi(v) \log \Lambda}{v^2}
\end{equation}
with the Coulomb logarithm $\log\Lambda$. The values for the Coulomb logarithm in the literature are $\log{\Lambda} = \left\{3, 10, \log \sqrt{m_1/m_2}\right\}$ \cite{Eda:2014kra,Yue:2019ozq,Kavanagh:2020cfn}. In this paper, we adopt the value used in \cite{Kavanagh:2020cfn}, $\log\Lambda = 
\log\sqrt{m_1/m_2}$. The factor $\xi(v)$ accounts for the fact that the particles in the dark matter halo are moving with different velocities relative to the orbiting object, first introduced in \cite{Kavanagh:2020cfn}. Dark matter particles only scatter and absorb momentum from the orbiting object if they are moving with a slower velocity compared to it. 

To calculate the density of particles moving slower than the orbital speed $v$, we can use \eqref{eq:rho_f}
\begin{equation}
\label{eq:rho_xi}
    \rho_\dm(r)\xi(v) = 4\pi \int_0^{v} v'^2 f\left(\Psi(r) - \frac{1}{2}v'^2 \right) dv'.
\end{equation}
Numerically, we find $\xi(v) \sim (v/v_{\text{orb}})^{3}$ for $v < v_{\text{max}}$, independent of radius. Only the circular orbital velocity $v_{\text{orb}}=\sqrt{\frac{m}{a}}$ changes with radius. This means that for circular orbits (as in \cite{Kavanagh:2020cfn}), where the secondary object always moves at $v=v_{\text{orb}}$, this can be approximated as a constant. In the above reference, a value of $\xi(v)\approx 0.58$ for a static halo with $\alpha=7/3$ has been calculated, but as we are looking at Keplerian orbits, the velocity of the orbiting object changes throughout one orbit. Therefore we cannot approximate it as a constant and instead need to include the phase space description into the differential equations.

\subsubsection{Orbital Evolution}
We are interested in the secular evolution of the orbital parameters $a(t),e(t)$ and the mass of the secondary object $m_2(t)$ under the backreaction of the dissipative forces. 

We can use \eqref{eq:E_orbit} to obtain
\begin{align}
    \pdv{E_{\text{orb}}}{a} =& \frac{m_2 m_1}{2a^2} \\
    \dv{a}{t} =& \dv{E_{\text{orb}}}{t}   / \pdv{E_{\text{orb}}}{a}. \label{eq:da_dt}
\end{align}
In a similar fashion, the evolution for $e$ can be derived from \eqref{eq:L_orbit} as
\begin{equation}
\label{eq:de_dt}
    \dv{e}{t} = -\frac{1-e^2}{2e} \left(\dv{E_{\text{orb}}}{t}/E_{\text{orb}} + 2\dv{L_{\text{orb}}}{t}/L_{\text{orb}} \right).
\end{equation}
By combining Eqs. \neqref{eq:da_dt} and \neqref{eq:de_dt} with Eqs. \neqref{eq:dEorb_dt} and \neqref{eq:dLorb_dt}, we obtain a system of differential equations that can be solved numerically.

\subsection{Gravitational Wave Signal}
The binary system emits GWs as a result of the change in the quadrupole moment. The system can be described with two polar angles $\iota$ and $\beta$. The inclination angle $\iota$ is given by the inclination of the plane of the orbit to the plane of the sky, while $\beta$ is the angle formed by the major axis and the direction of the observer in the orbital plane (see for example Fig. 1 in \cite{Moreno-Garado:1995msd}). The gravitational strain calculation of the two polarizations for a Keplerian orbit yields \cite{Martel:1999tm}

\begin{widetext}

\begin{flalign}
    h_+ = & - \frac{m\mu}{p D_L} \Bigg[  \left( 2 \cos(2\phi - 2\beta) + \frac{5}{2}e \cos(\phi - 2\beta) + \frac{1}{2}e \cos(3\phi - 2\beta) + e^2 \cos(2\beta)   \right) (1+\cos^2\iota ) &&\nn\\
    & \quad\quad\quad + (e\cos\phi + e^2) \sin^2\iota      \Bigg]  &&\\
    h_{\cross}  = & - \frac{m\mu}{p D_L} \Bigg[ 4\sin(2\phi-2\beta) + 5e\sin(\phi-2\beta) + e\sin(3\phi-2\beta) - 2e^2 \sin(2\beta)  \Bigg] \cos(\iota) &&
\end{flalign}
\end{widetext} 
for a system at luminosity distance $D_{\text{L}}$ and  semilatus rectum $p=a(1-e^2)$.

\subsubsection{Harmonic Decomposition}
The GW signal can be decomposed into the harmonics of the mean orbital frequency as follows
\cite{Moreno-Garado:1995msd, Moore:2018kvz}
\begin{equation}
\label{eq:h_harmonics}
    h_{+,\cross} = \mathcal{A} \sum_{n=1}^{\infty} \left(C_{+,\cross}^{(n)} \cos(n\ell) + S_{+,\cross}^{(n)} \sin(n\ell)\right)
\end{equation}
with the mean anomaly 
\begin{equation}
    \ell(t) = \int^t \mathrm{d}t\,  2\pi \F 
\end{equation}
 and the amplitude
\begin{equation}
    \mathcal{A} = -\frac{\mchirp}{D_L} (2\pi\mchirp \F)^{2/3}.
\end{equation}
that depends on the chirp mass $\mchirp = \mu^{3/5}m^{2/5}$ of the system.

The coefficients $C_{+,\cross}^{(n)}, S_{+,\cross}^{(n)}$ can be obtained by using \eqref{eq:Kepler_r} and the Fourier-Bessel expansion of the orbital motion, as described in \cite{Moreno-Garado:1995msd, PhysRevD.80.084001, Moore:2018kvz}. This gives the coefficients \cite{Chandramouli:2021kts}
\begin{widetext}
\begin{subequations}
\begin{align}
    C_{+}^{(n)}  = &  \Big[ 2s_\iota^2 J_n(ne) + \frac{2}{e^2}(1 + c_\iota^2)c_{2\beta} \left( 
        (e^2 - 2)J_n(ne) + ne(1-e^2) (J_{n-1}(ne) - J_{n+1}(ne))\right) \Big],  \\
    S_{+}^{(n)}  = & - \frac{2}{e^2} \sqrt{1-e^2} (1+c_\iota^2)s_{2\beta} \Big[ -2(1-e^2)nJ_n(ne) + e(J_{n-1}(ne) - J_{n+1}(ne)) \Big], \\
    C_{\cross}^{(n)}  = & - \frac{4}{e^2} c_\iota s_{2\beta} \Big[ (2-e^2)J_n(ne) + ne(1-e^2)(J_{n-1}(ne) - J_{n+1}(ne))  \Big], \\
    S_{\cross}^{(n)}  = & - \frac{4}{e^2} \sqrt{1-e^2} c_\iota c_{2\beta} \Big[ -2(1-e^2)nJ_n(ne) + e(J_{n-1}(ne) - J_{n+1}(ne))\Big],
\end{align}  
\end{subequations}
\end{widetext}
where the $J_n$ are Bessel functions of the first kind, and $c_{\beta}=\cos (\beta)$, $s_{\beta}=\sin(\beta)$. These equations are valid for any eccentricity $e$.

\subsubsection{Stationary Phase Approximation}
To calculate the signal in the frequency domain, the \textit{stationary phase approximation} is used, which is described in Appendix A of \cite{Moore:2018kvz}. As the amplitude only varies slowly over time, the Fourier transform integrates over rapidly oscillating sinusoidal terms in \eqref{eq:h_harmonics}, which are negligibly small, except when the \textit{stationary phase condition } 
\begin{equation}
    n\F(t^*_n) = f
\end{equation}
for a given time $t^*_n$ is fulfilled. This suggests that an eccentric binary emits at all integer multiples of the mean orbital frequency. The stationary phase condition gives a mapping between the time $t_n^*$ and frequency of the $n$-th harmonic. All harmonics are emitted at any given time, so there is no one-to-one correspondence between observed frequency and time, as in the circular case.

With the stationary phase approximation, the Fourier transform of the signal can be obtained, which reads for a given harmonic \cite{Moore:2018kvz, Chandramouli:2021kts}
\begin{align}
    \tilde{h}_{+,\cross}^{(n)}(f) =& - \frac{\mchirp}{2D_L} \frac{(2\pi \mchirp \F(t^*_{n}))^{2/3}}{\sqrt{n \dot{\F}(t^*_{n})} } \nn \\
     &   \times\Big[C_{+,\cross}^{(n)}(t^*_{n}) + i S_{+,\cross}^{(n)}(t^*_{n}) \Big] e^{i\psi_n},  
\end{align}
where
\begin{equation}
    n\F_0 < f < n\F_\isco.
\end{equation}
Since the evolution takes place over a finite time, only a finite range of frequencies are emitted for a given harmonic. Therefore, the frequency ranges from some initial frequency of the system $\mathcal{F}_0$ to the final frequency of the Last Stable Orbit (LSO), which can be approximated by the Innermost Stable Circular Orbit (ISCO) for low eccentricities. 

The phase of the harmonic is given by 
\begin{equation}
    \psi_n = 2\pi ft^*_n - n \ell - \frac{\pi}{4}.
\end{equation}
This reduces to Equation (25b) of \cite{Eda:2014kra} for the $n=2$ case.

\subsubsection{Dephasing}
To observe the effect the dark matter halo has on the evolution, we can look at the \textit{dephasing}. To this end, we compare the number of GW cycles completed in the cases with and without dark matter present, following \cite{Kavanagh:2020cfn}. We can do this for each harmonic individually between some initial time $t_\text{i}$ and final time $t_\text{f}$  with
\begin{equation}
    N^{(n)}(t_\text{f}, t_\text{i}) = n \int_{t_\text{i}}^{t_\text{f}} \F(t)\mathrm{d}t.
\end{equation}
Setting $t_\text{f} = t_\text{c}$ as the time of coalescence, we obtain
\begin{equation}
    \Delta N^{(n)}(t)= N^{(n)}_\text{vacuum}(t_\text{c}, t) - N^{(n)}_\text{DM}(t_\text{c}, t).
\end{equation}
The dephasing effect is stronger for larger harmonics, as they complete more phases in the same time span. Unfortunately, while the system emits at all harmonics, their contribution will not necessarily be observable. For low eccentricities, the system emits primarily in the $n=2$ harmonic, like in the circular case. For higher eccentricities, the system generally emits at higher harmonics, see, for example, Fig. 2 in \cite{Moore:2018kvz}. As the eccentricity evolves, the observable harmonics can change over time, making the dephasing effect difficult to track. To assess detectability, we need to look at the detector sensitivity.

\subsection{Detector Sensitivity}
To assess detectability, we consider the dimensionless characteristic strain of the GW signal  \cite{Moore:2014lga}
\begin{equation}
    [h_c(f)]^2 = 4f^2\abs{\tilde{h}(f)}^2 .
\end{equation}
This needs to be compared to the noise amplitude
\begin{equation}
    [h_n(f)]^2 = fS_n(f),
\end{equation}
where $S_n(f)$ is the Power Spectral Density (PSD) function of the noise of the detector. For LISA, we use the PSD function given by Eq. (13) of \cite{Robson:2018ifk}. The signal to noise ratio is then expressed as
\begin{equation}
    \varrho^2 = \int_{-\infty}^{\infty} d\log(f) \abs{\frac{h_c(f)}{h_n(f)}}^2
\end{equation}
Thus, a plot of the characteristic strain and the noise amplitude allows one to easily assess the detectability of a given signal.

%% file: results.tex
In this section we present the results from the numerical integration of the system of differential equations. The equations have been implemented in \textssc{python} and numerically evolved and evaluated. The code is publicly available and can be found at: \url{https://github.com/DMGW-Goethe/imripy}. 

\subsection{Inclusion of $\xi(v)$}
Let us start by exploring the results of the inclusion of the term $\xi(v)$ (\eqref{eq:rho_xi}) in the dynamical friction force. 

\begin{figure}
    \centering
    \includegraphics[width=1\textwidth]{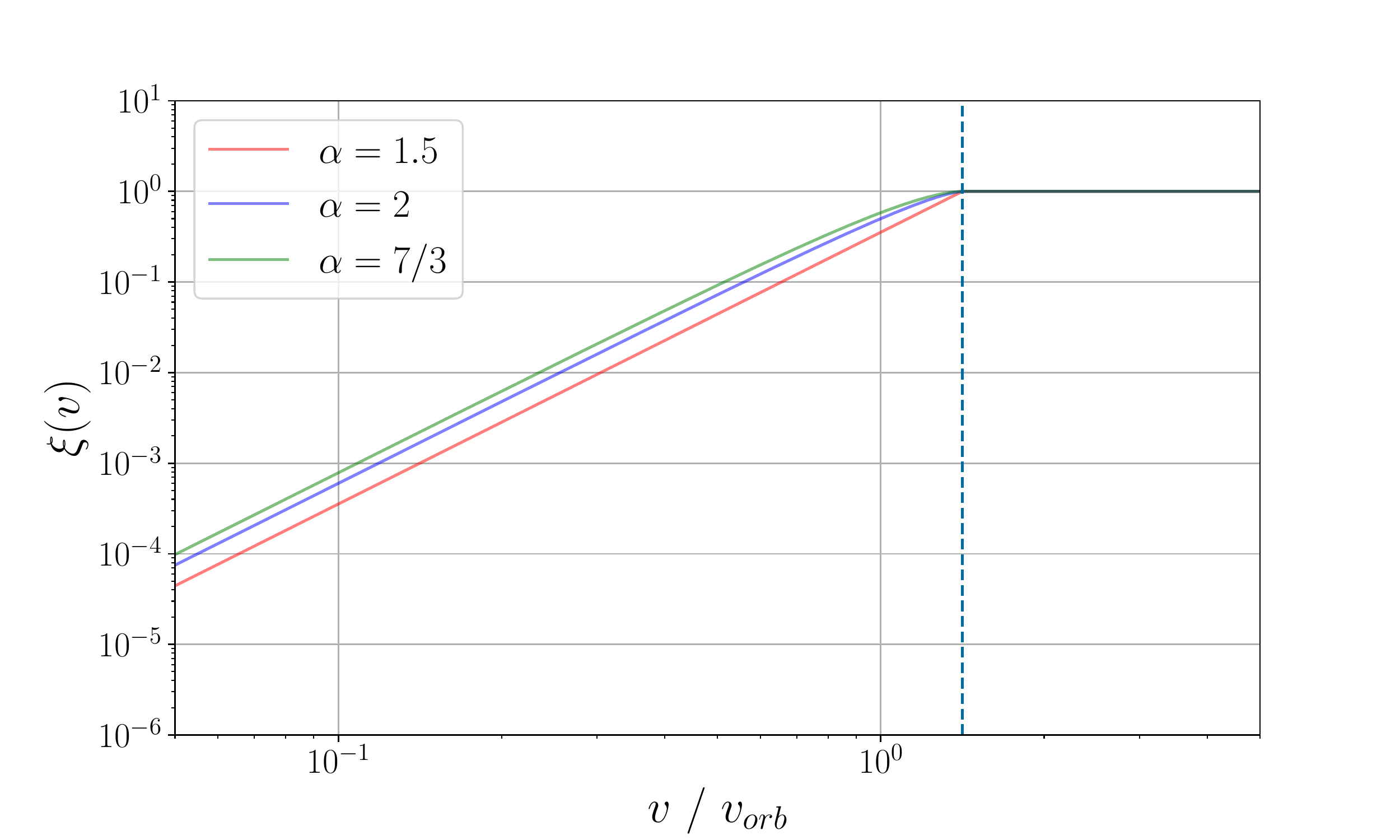}
    \caption{The phase space factor, which selects only the particles that are moving slower than a given speed $v$ from the dark matter halo. This is for a dark matter halo with $\{m_1, \rho_6\} = \{10^3 \Msun, 5.448\times 10^{15}\Msun$/\pc$^3 \}$. The dotted line marks $v_\text{max}=\sqrt{2}\,v_\text{orb}$, the escape velocity at the given orbital radius. All particles bound in the dark matter halo move more slowly than this, thus $\xi(v)$ caps at $1$. At smaller velocities, a power law behavior can be seen with $\xi(v)\propto v^3$ for the different $\alpha_\sp = \{1.5, 2, 7/3 \}$. }
    \label{fig:xi}
\end{figure}

A plot of $\xi(v)$ is shown in \figref{fig:xi} for halos with $\alpha_\sp =  \{1.5, 2, 7/3 \}$. At the orbital velocity it can be seen that $\xi(v_\text{orb})\approx 0.58$ for $\alpha_\sp=7/3$, as claimed in \cite{Kavanagh:2020cfn}. The dotted line marks $v_\text{max} = \sqrt{2}\,v_\text{orb}$, which is the escape velocity at the given radius, above which there can be no orbiting particle in the dark matter halo. Below $v_\text{orb}$, a power law behavior can be seen with $\xi(v)\propto v^3$, as claimed in the previous section.

\begin{figure}
    \centering
    \includegraphics[width=1.1\textwidth]{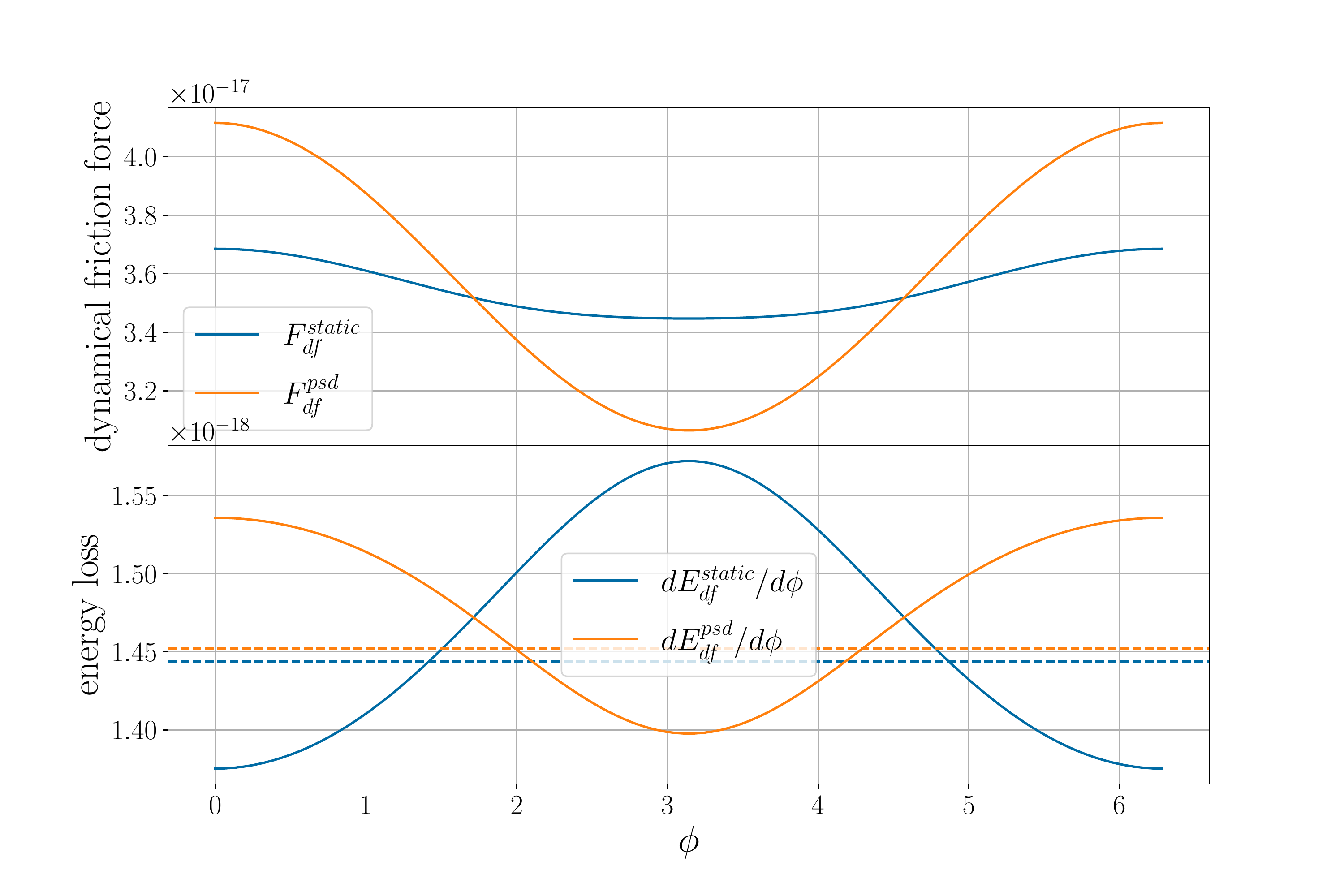}
    \caption{The dynamical friction force and energy loss over one eccentric orbit. This is done for both the case with $\xi(v)$ (\textit{psd}) and without (\textit{static}). The parameters of the halo and orbit are $\{m_1, m_2, \rho_6, \alpha_\sp, a_0, e_0 \} = \{10^3 \Msun, 1 \,\Msun, 5.448\times 10^{15}\Msun$/\pc$^3, 7/3, 100 \;r_\isco, 0.1 \}$. The change of shape can be observed in the energy loss. The dotted lines represent the orbital average.  }
    \label{fig:F_df}
\end{figure}

The inclusion of $\xi(v)$ has a drastic effect on the nature of the dynamical friction force on a Keplerian orbit. This is shown in \figref{fig:F_df}, where the friction force (\eqref{eq:F_df}) is plotted over one orbital revolution, characterized by the true anomaly $\phi$. This is done including and excluding the $\xi(v)$ term, referred to as phase space distribution (\textit{psd}) and \textit{static} cases, respectively. Along with it, the energy loss as given by the integrand of \eqref{eq:avgdEdT} is shown. It can be seen that the inclusion of $\xi(v)$ changes the behavior of the energy loss over one orbit. Following \eqref{eq:Kepler_r}, $\phi = 0$ is the periapsis, the closest point in the orbit, and $\phi=\pi$ the apoapsis, the furthest point in the orbit. In the \textit{static} case, the energy loss is strongest at the apoapsis, which leads to the eccentrification of the orbit as observed by \cite{Yue:2019ozq}. In the  \textit{psd} case, the energy loss is strongest at the periapsis, which leads to circularization.

Intuitively, the difference can be explained by the fact that with $\xi(v)$, the secondary object only scatters with DM particles that are moving slower than it. Further out in the orbit, the object is moving slower than it would on a circular orbit. This means that there are comparatively less interactions further out in the orbit. At the same time, there are more interactions on the inner parts of the orbit, where the object is moving faster than in the circular case. Therefore, the force weakens further out in the orbit and is stronger on the inner parts of the orbit, compared to the \textit{static} case.

\begin{figure}
    \centering
    \includegraphics[width=1.1\textwidth]{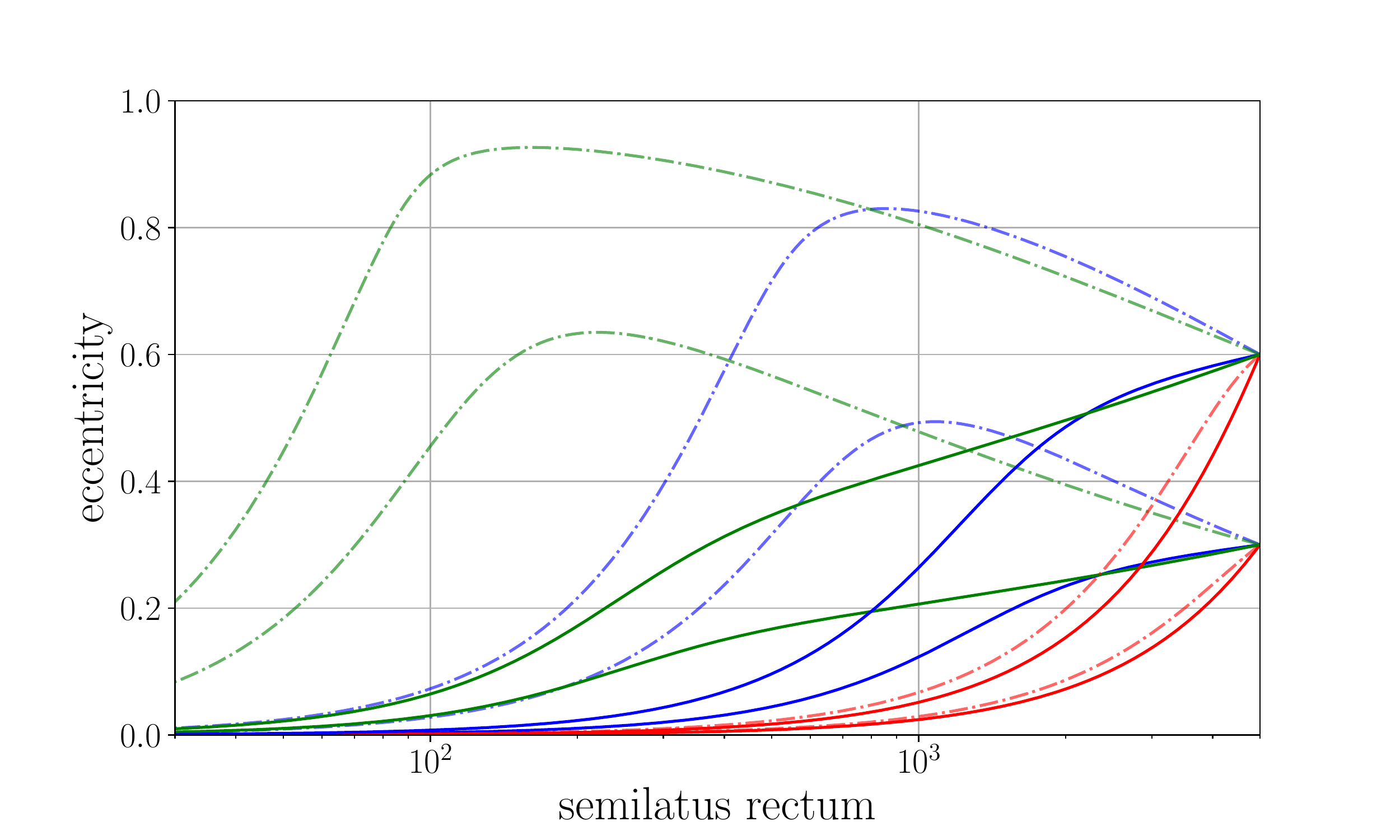}
    \caption{The evolution of the eccentricity $e$ as a function of semilatus rectum $p$, modeled after Fig. 1b in \cite{Yue:2019ozq}. The semilatus rectum decreases monotonically with time, so the temporal evolution is from right to left. The red, blue, green lines correspond to a spike power law index of $\alpha_\text{spike}=\{1.5, 2, 7/3\}$ respectively. The dashed lines represent the evolution with $\log \Lambda = 10$ and $\xi(v)\equiv 1$, while the solid lines correspond to $\log \Lambda=\log\sqrt{m_1/m_2}$, and $\xi(v)$ as given by \eqref{eq:rho_xi}. The initial semilatus rectum is $\tilde{p}_0=5000$ and initial eccentricity is $e_0=\{0.3, 0.6\}$. The other parameters of the system are as in \cite{Yue:2019ozq}: $\{m_1, m_2, \rho_\text{spike}, r_\text{spike} \} = \{10^3 \Msun, 10 \,\Msun, 226\,\Msun/\pc^3, 0.54\,\pc \}$. }
    \label{fig:xi_pe}
\end{figure}

For comparison, we can look at the systems analysed in \cite{Yue:2019ozq}. The system's evolution is plotted in \figref{fig:xi_pe}, which is modeled after Fig. 1b of \cite{Yue:2019ozq}, and has the same system parameters $\{m_1, m_2, \rho_\text{spike}, r_\text{spike} \} = \{10^3 \Msun, 10 \,\Msun, 226\,\Msun/\pc^3, 0.54\,\pc \}$. We have modeled three power laws corresponding to $\alpha_\sp = \{ 1.5, 2, 7/3\}$. The dotted lines represent the evolution with their model, $\xi(v)\equiv 1$ and $\log \Lambda =10$. The solid lines are the results of the evolution with the model from section \ref{sec:equations}. 

The evolution is presented in a plot of eccentricity versus dimensionless semilatus rectum $\tilde{p} = a(1-e^2)/m_1$, for purposes of comparison. The temporal evolution is from right to left in the plot, as the semilatus rectum decreases during the inspiral. Two effects can be observed here. First, there is no eccentrification of the orbit, due to the inclusion of $\xi(v)$, as expected from the previous paragraph. We will further analyze this in section \ref{sec:analysis}. Second, there are two regimes that can be observed in both cases. Initially (to the right of the plot), the dynamical friction dominates the energy loss. Later in the inspiral, the GW emission dominates the energy loss. This leads to a stronger circularization of the orbit. For the case with $\xi(v)$, this can be seen by a change in the slope of the eccentricity. This change happens earlier than in the case of \cite{Yue:2019ozq}, which is due to a smaller $\log \Lambda$. Notice that \cite{Yue:2019ozq} uses $\log \Lambda = 10$, while we use $\log \Lambda= \log \sqrt{m_1/m_2}= \log 10 \approx 2.3 $.  This results in an earlier dominance of the GW emission loss over the dynamical friction effects.

\begin{figure}
    \centering
    \includegraphics[width=1.1\textwidth]{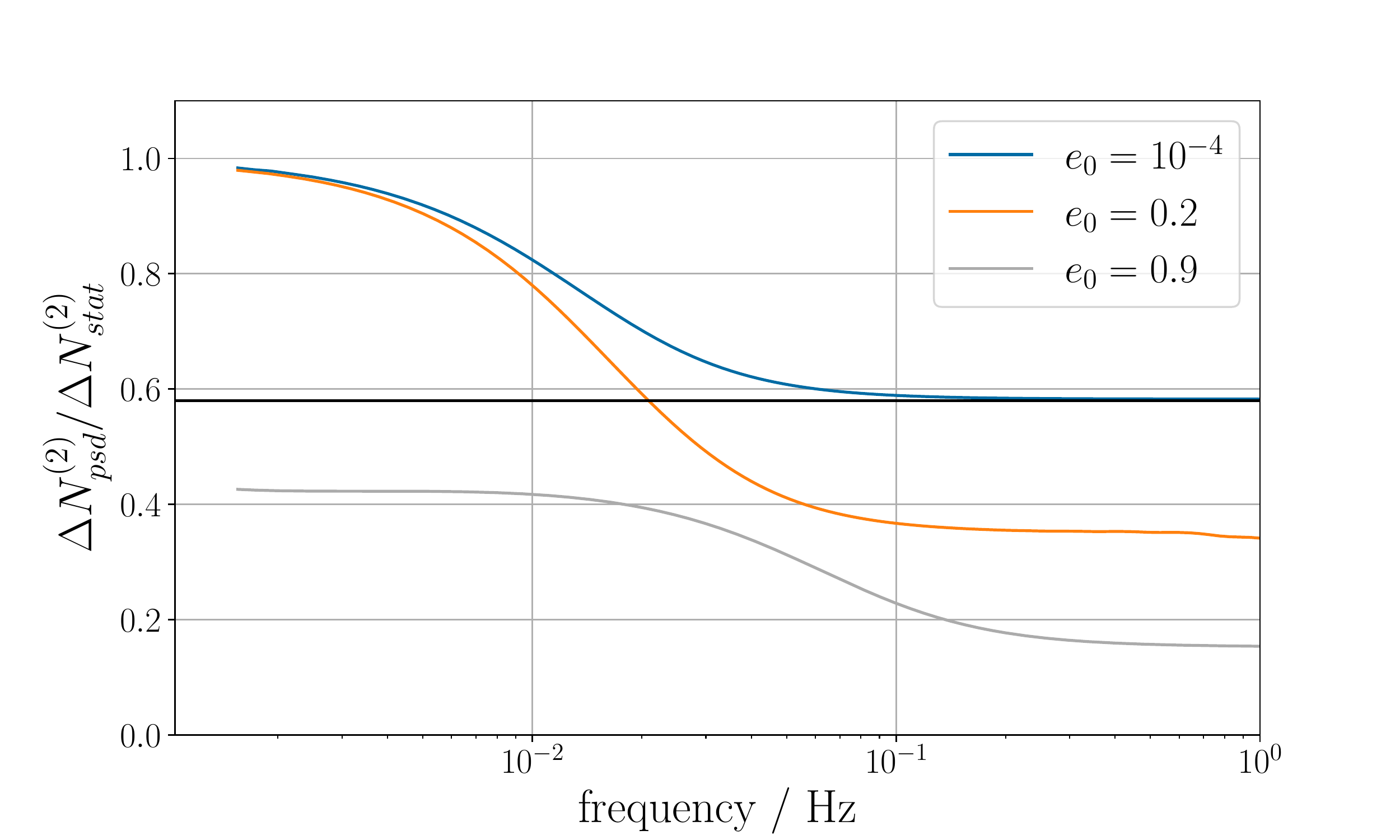}
    \caption{The relative dephasing $\Delta N^{(2)}_\text{dyn}/\Delta N^{(2)}_\text{stat}$ for the \textit{static} ($\xi(v)\equiv 1$) and \textit{psd} case for different initial eccentricities $e_0= \{ 10^{-4} , 0.2, 0.9 \}$. The solid black line is at the value of $0.58$, which is quoted by \cite{Kavanagh:2020cfn} to be the relative reduction of the dynamical friction force in the \textit{psd} case (for $\alpha_\sp = 7/3$). The other parameters of the system are $\{m_1, m_2, \rho_\sp, r_\sp, a_0\} = \{10^3\Msun, 1\,\Msun, 226 \,\Msun/\pc^3, 0.54 \,\pc, 200 \;r_\isco \}$. }
    \label{fig:xi_dephasing}
\end{figure}

Finally, we want to compare the dephasing effects that the inclusion of $\xi(v)$ brings about. This is shown in \figref{fig:xi_dephasing}. The dephasing in the second harmonic, $\Delta N^{(2)}$, can be seen to be dependant on the initial eccentricity. For the almost circular case, $e_0 = 10^{-4}$, the ratio between the amount of dephasing can be seen to approach $0.58$, which is the value presented in \cite{Kavanagh:2020cfn} as the reduction factor of the dynamical friction strength in the circular case. 
For higher eccentricities, the relative impact in the \textit{psd} case is stronger (as it gets further away from $1$). This can be explained by the fact that GW emission is stronger for such higher eccentricities. Thus, as the eccentricity is higher in the \textit{static} case, GW emission dominates earlier and speeds up the inspiral, compared to the \textit{psd} case.

\FloatBarrier % this forces the figures to be placed before this barrier
\subsection{Spike Profiles}
\begin{figure}
    \centering
    \includegraphics[width=1\textwidth]{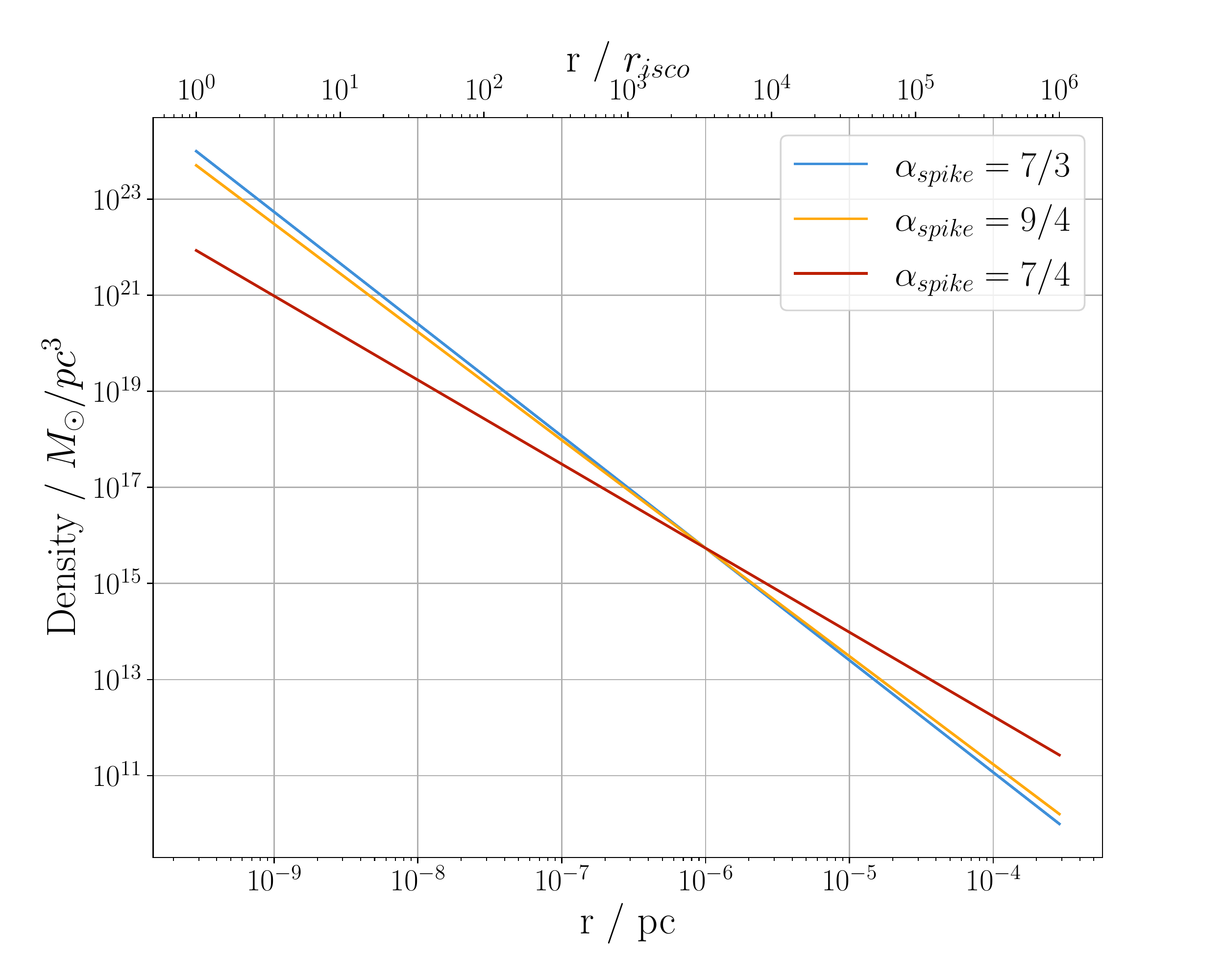}
    \caption{The dark matter density profile for the different power laws $\alpha_\sp = \{7/3, 9/4, 7/4 \}$ with the density parameter $\rho_6=5.448\times 10^{15}\Msun/\pc^3$. }
    \label{fig:rho_dm}
\end{figure}

In this section we focus on the \textit{psd} case with the inclusion of $\xi(v)$ and explore different power law indices. The values considered here are $\alpha_\sp = \{7/3, 9/4, 7/4 \}$. 

\begin{figure}
     \centering
     \hspace{0.5cm}\includegraphics[width=1.1\textwidth]{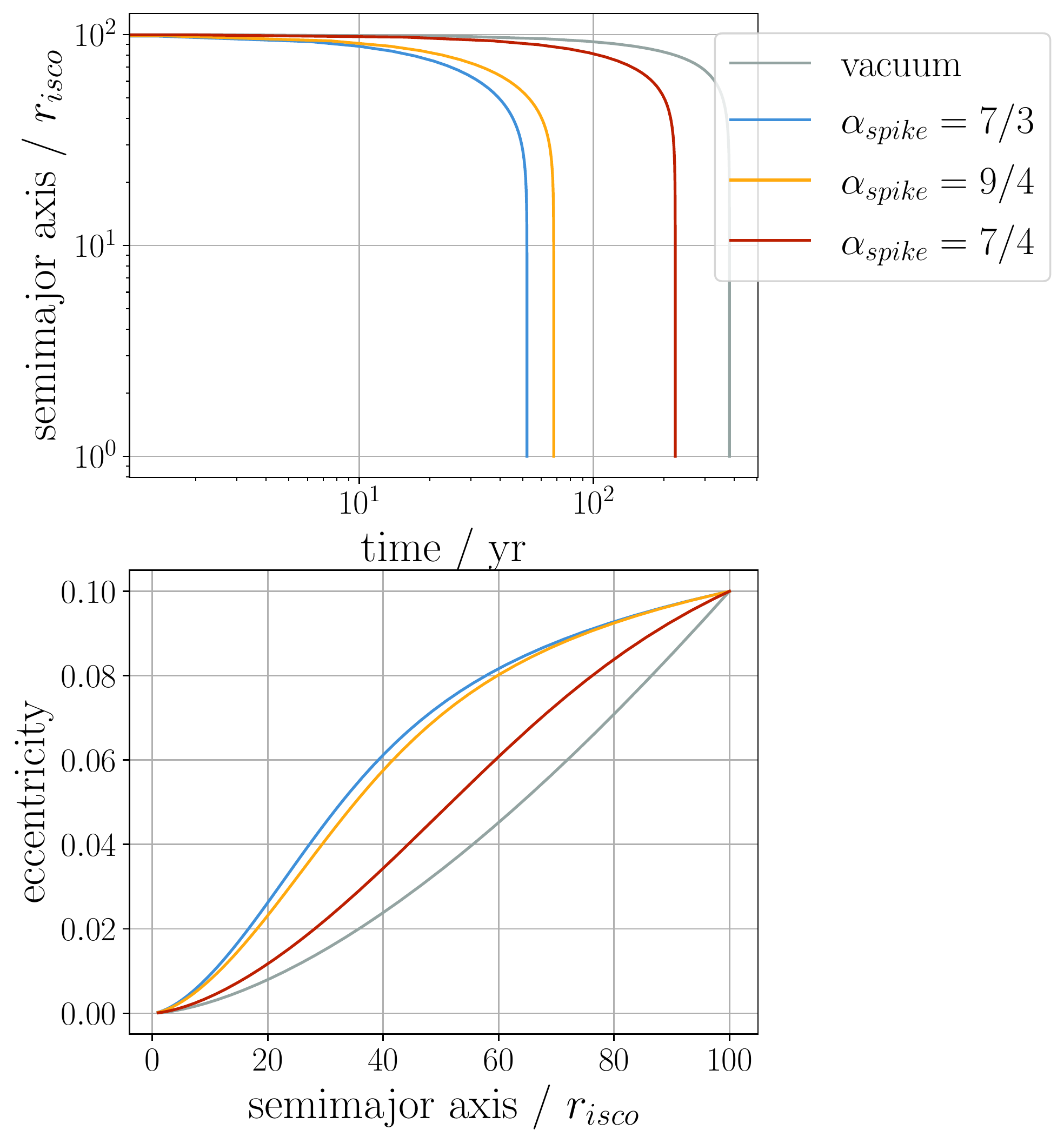}\\
     \includegraphics[width=\textwidth]{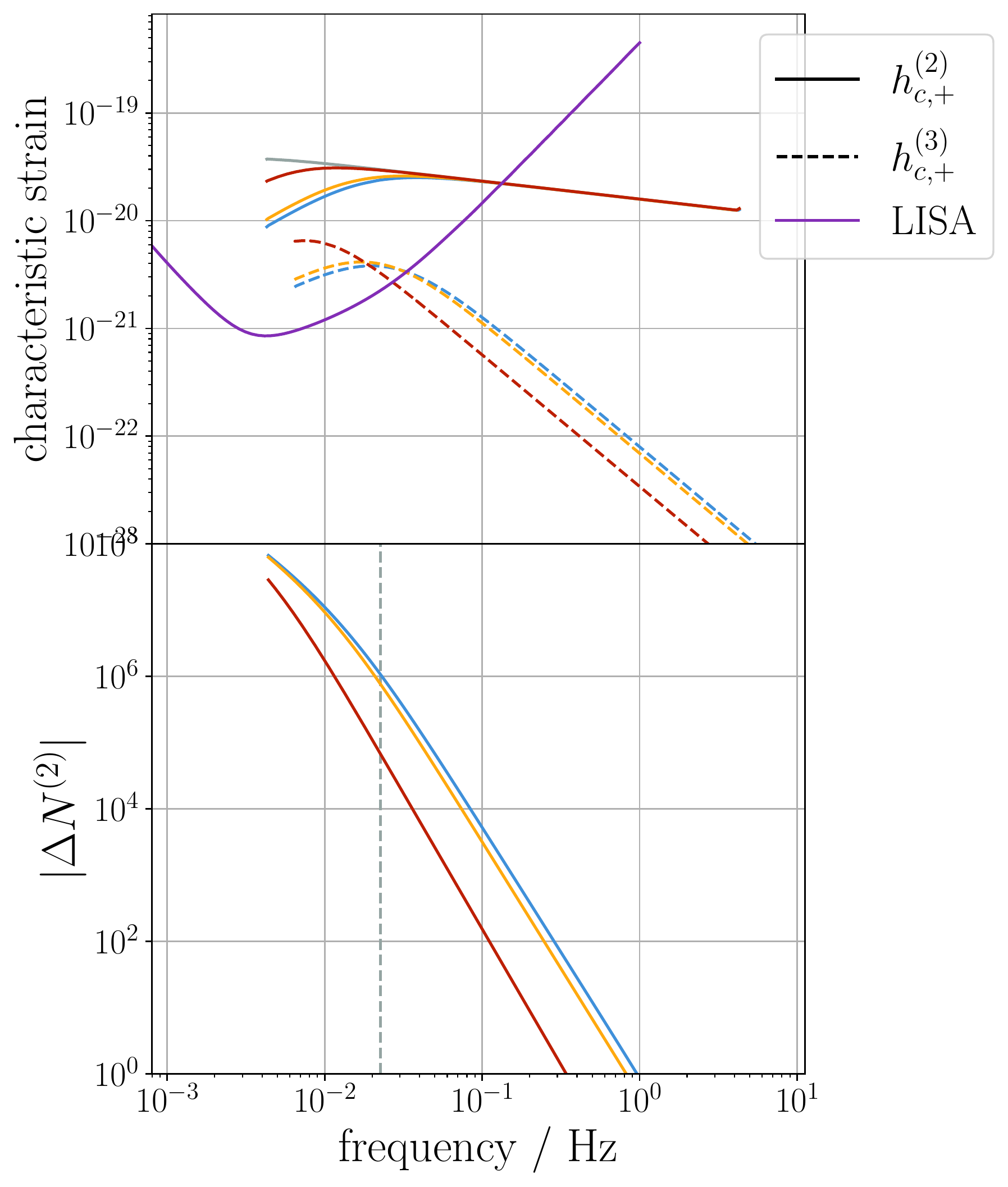}
     \caption{\textbf{Top}: The evolution of the semimajor axis $a$. Below, the evolution of the eccentricity $e$, depending on the semimajor axis. \textbf{Bottom}: The characteristic strain of the GW signal $h^{(2)}_c$(solid), $h^{(3)}_c$(dashed) compared to the LISA sensitivity, and the dephasing $\Delta N^{(2)}$. The system parameters are $\{D_L, m_1,m_2, \rho_6, \alpha_\sp, a_0, e_0\} = \{500\,\text{Mpc}, 10^3 \Msun, 1\,\Msun, 5.448\times 10^{15}\Msun/\pc^3, 7/3, 100\;r_\isco, 0.1 \}$.  }
     \label{fig:evolution_gwsignal}
 \end{figure}

As an example, we consider the system analyzed in \cite{Kavanagh:2020cfn} with $m_1=10^3\Msun$, $m_2=1\,\Msun$, $\rho_6=5.448\times 10^{15}\Msun$/\pc$^3$, at a luminosity distance of $d_L=100\,$Mpc. An initial semimajor axis is chosen with $a_0=100\;r_\text\isco$, such that the system inspirals on the order of $\sim 10$ years, and the initial eccentricity is set to $e_0 = 0.1$. The dark matter density is plotted in \figref{fig:rho_dm} and the results of the numerical integration are shown in \figref{fig:evolution_gwsignal}. 

There, the three different power law spikes are plotted, along with the case without dark matter. In the evolution of the semimajor axis $a$ over time it can be seen that the inspiral time is significantly shortened, compared to the case with no dark matter. The effect is stronger for larger power laws, because they result in higher dark matter densities in the region of inspiral. 

The evolution of the eccentricity in relation to the semimajor axis is plotted as well. A similar behavior is observed here, the larger power law spikes have a stronger effect on the evolution. As expected, the dynamical friction circularizes the orbit, but at a different rate than GW emission loss. Early on in the evolution, the dynamical friction effects dominate and the eccentricity is slowly reduced, compared to when the GW emission loss dominates later on.

This can also be seen in the GW signal of the system. The characteristic strain of the second and third harmonic are shown for the cases, and they behave differently for different power laws. Both harmonics are in the observable band of LISA for the given luminosity distance. The initial rising slopes are due to the region where dynamical friction loss dominates over the GW emission loss and the typical $h_c^{(2)} \sim f\tilde{h}^{(2)} \sim f^{-1/6}$\cite{Cutler:1994ys} dependency is only later recovered. Intuitively, the system spends less time emitting at low frequencies and thus the spectrum is lower in Fourier space. The third harmonic can be seen to decay away faster as the system is circularizing. The rate of circularization is observable in the ratio of the second and third harmonic. The rate of circularization depends on the local dark matter density, but also on the power law index, as the next section will show. 

Finally, the dephasing effects are shown for the second harmonic $\Delta N^{(2)}$, which is the dominant one in this case. The dephasing mostly depends on the local dark matter density, which is shown in the overall behavior of $\Delta N^{(2)}$. The cycle difference is shown to be around $10^5 \sim 10^6$ at $f\sim 2\times10^{-2}$, where the system has 5 years left to inspiral, as seen in \cite{Kavanagh:2020cfn}.

 \subsubsection{Varying Initial Eccentricity}
 
 \begin{figure*}
    \includegraphics[width=\textwidth]{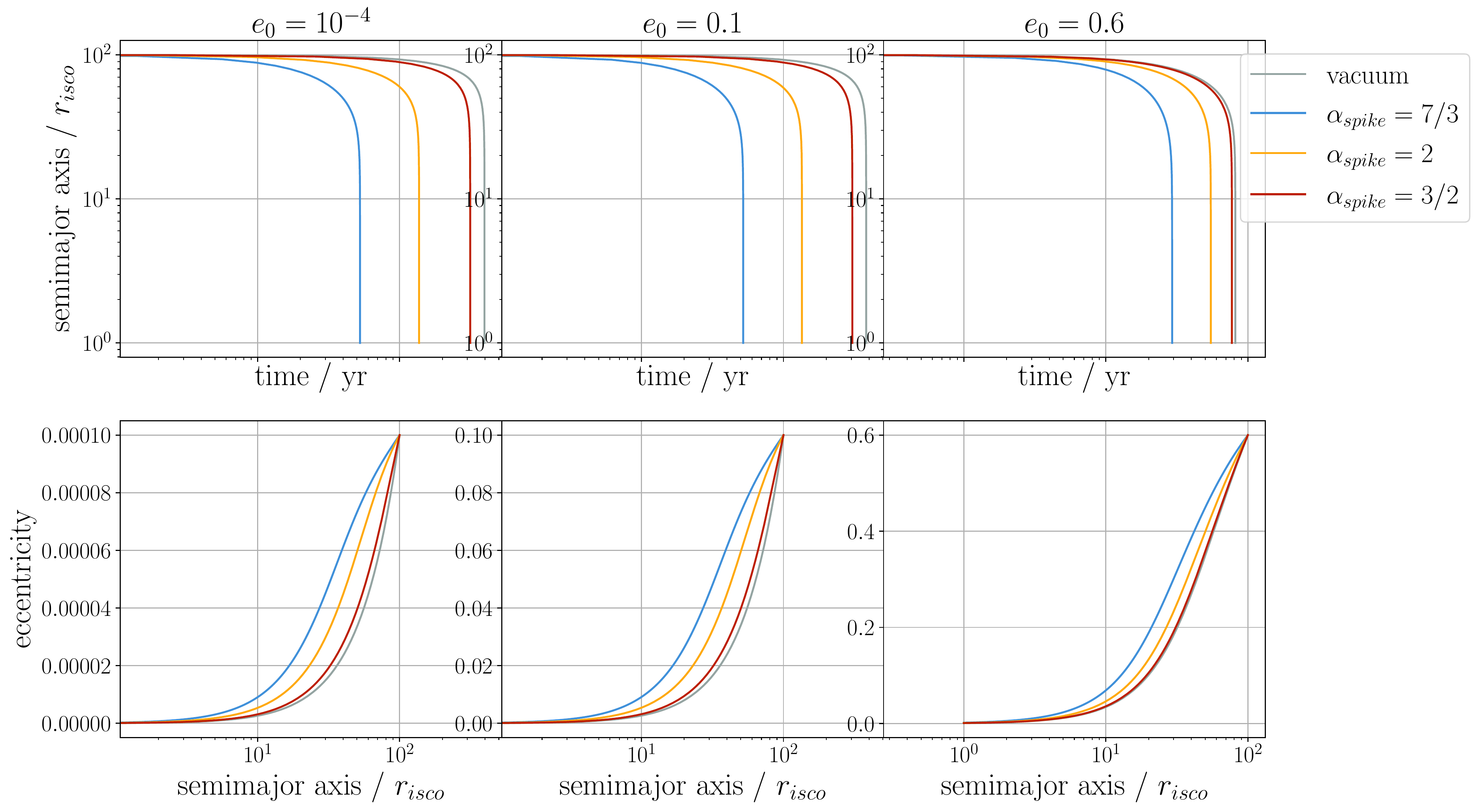}\\
    \hspace{-0.85cm}\includegraphics[width=0.935\textwidth]{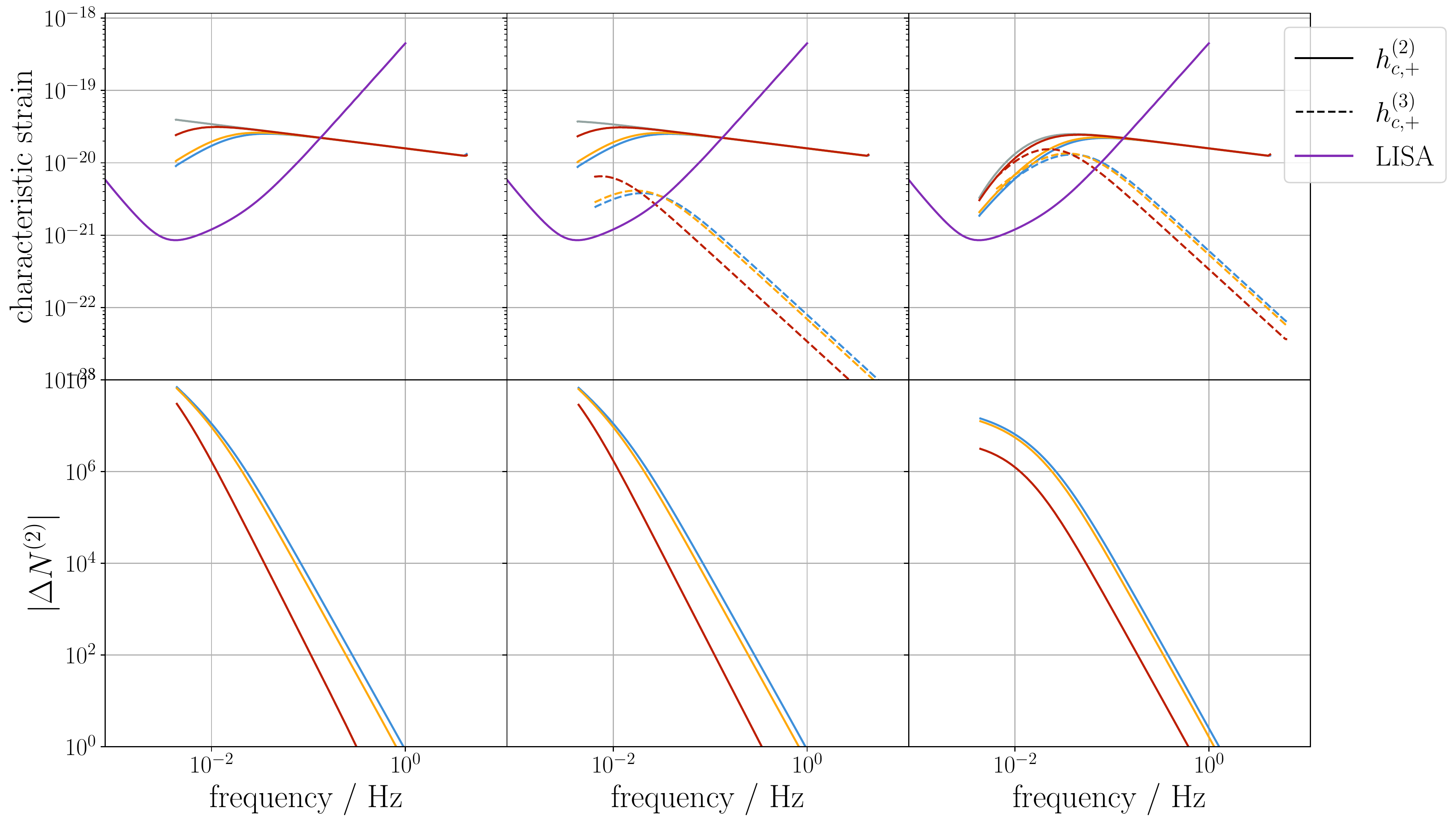}
    \caption{\textbf{Top}: The eccentricity evolution depending on the semimajor axis. The behavior is consistent for different initial eccentricities. \textbf{Bottom}: The characteristic strain of the GW signal $h^{(2)}_c$(solid), $h^{(3)}_c$(dashed) compared to the LISA sensitivity, and the dephasing $\Delta N^{(2)}$ for varying initial eccentricity $e_0=\{10^{-4}, 0.1, 0.6\}$. 
    The parameters of the system are $\{D_L, m_1,m_2, \rho_6, \alpha_\sp, a_0\} = \{ 500\text{Mpc}, 10^3 \,\Msun, 1\,\Msun, 5.448\times 10^{15}\Msun/\pc^3, 7/3, 100\; r_\isco \}$. 
    \label{fig:evolution_e0}}
\end{figure*}
 
For varying initial eccentricity $e_0$, we show the evolution for $e_0 = \{10^{-4}, 0.1, 0.6\}$ in \figref{fig:evolution_e0}. It can be seen that the evolution of the eccentricity is qualitatively similar in between the cases. Initially, there is a phase dominated by dynamical friction effects with slow circularization, later there is a phase dominated by GW emission loss, with faster circularization. This means that the system spends a significant amount of time close to its initial eccentricity. Therefore, real systems could be in principle observed with intermediate eccentricities. 

The characteristic strain shows the interplay of eccentricity and harmonics. The larger the eccentricity, the weaker the second and the stronger the third harmonic. For $e_0=0.6$, they are on equal grounds. For the highest eccentricities, $h_c^{(1)}$ is expected to be the dominant one \cite{Moore:2018kvz}. It should be easier to observe the dephasing effects for higher harmonics, since $\Delta N^{(n)} \sim \frac{n}{2}\Delta N^{(2)}$, while these are observable. This makes systems at intermediate eccentricities optimal to observe the dephasing.

It can be seen that the eccentricity of the system does not have a strong influence on the amount of dephasing $\Delta N^{(2)}$. Only for higher eccentricity, the overall dephasing effects are smaller, because the system inspirals faster for higher eccentricity. This was already observed in \figref{fig:xi_dephasing}.

\subsubsection{Varying Dark Matter Density}
\begin{figure*}
    \includegraphics[width=\textwidth]{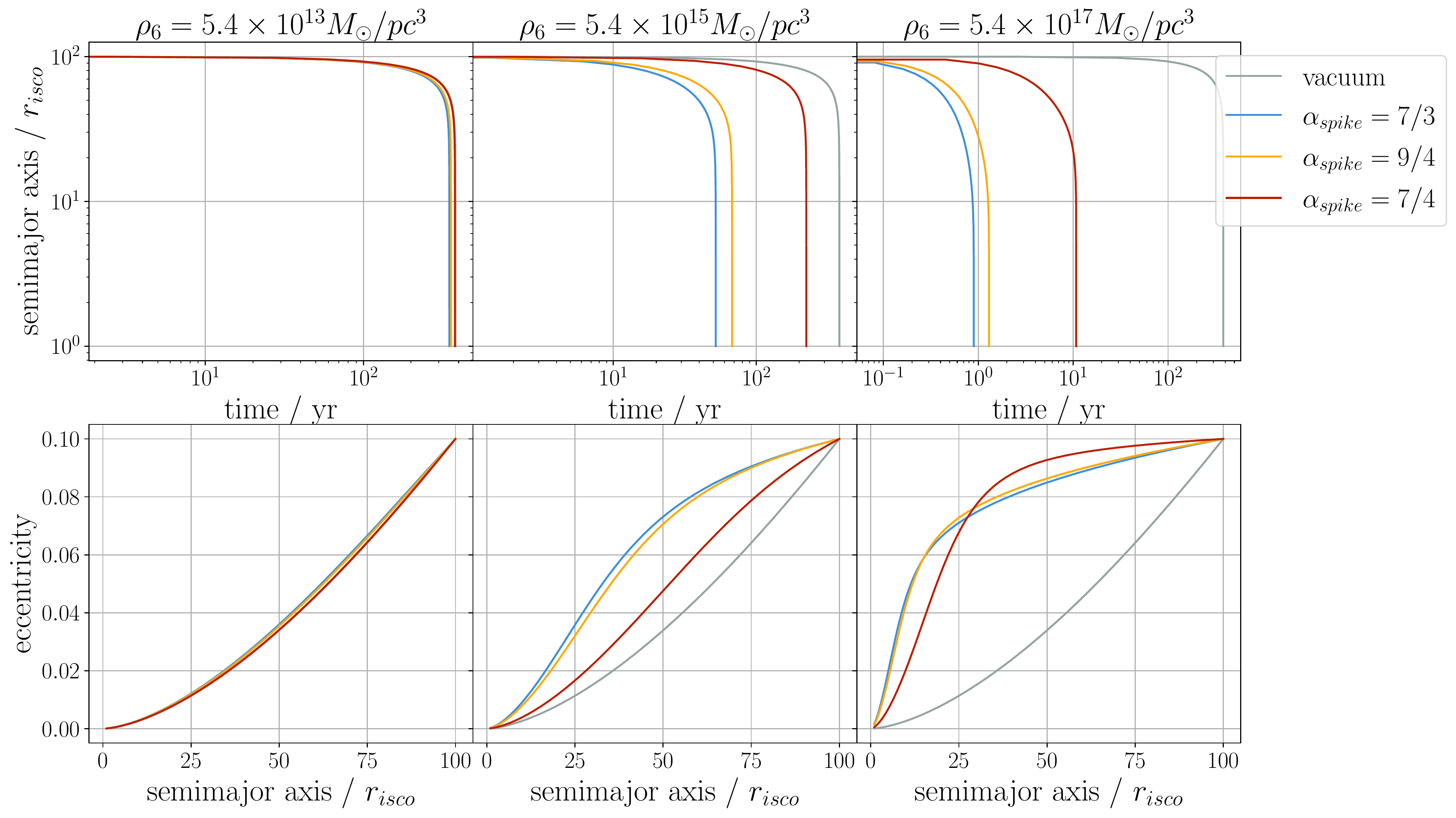}\\
    \hspace{-0.95cm}\includegraphics[width=0.955\textwidth]{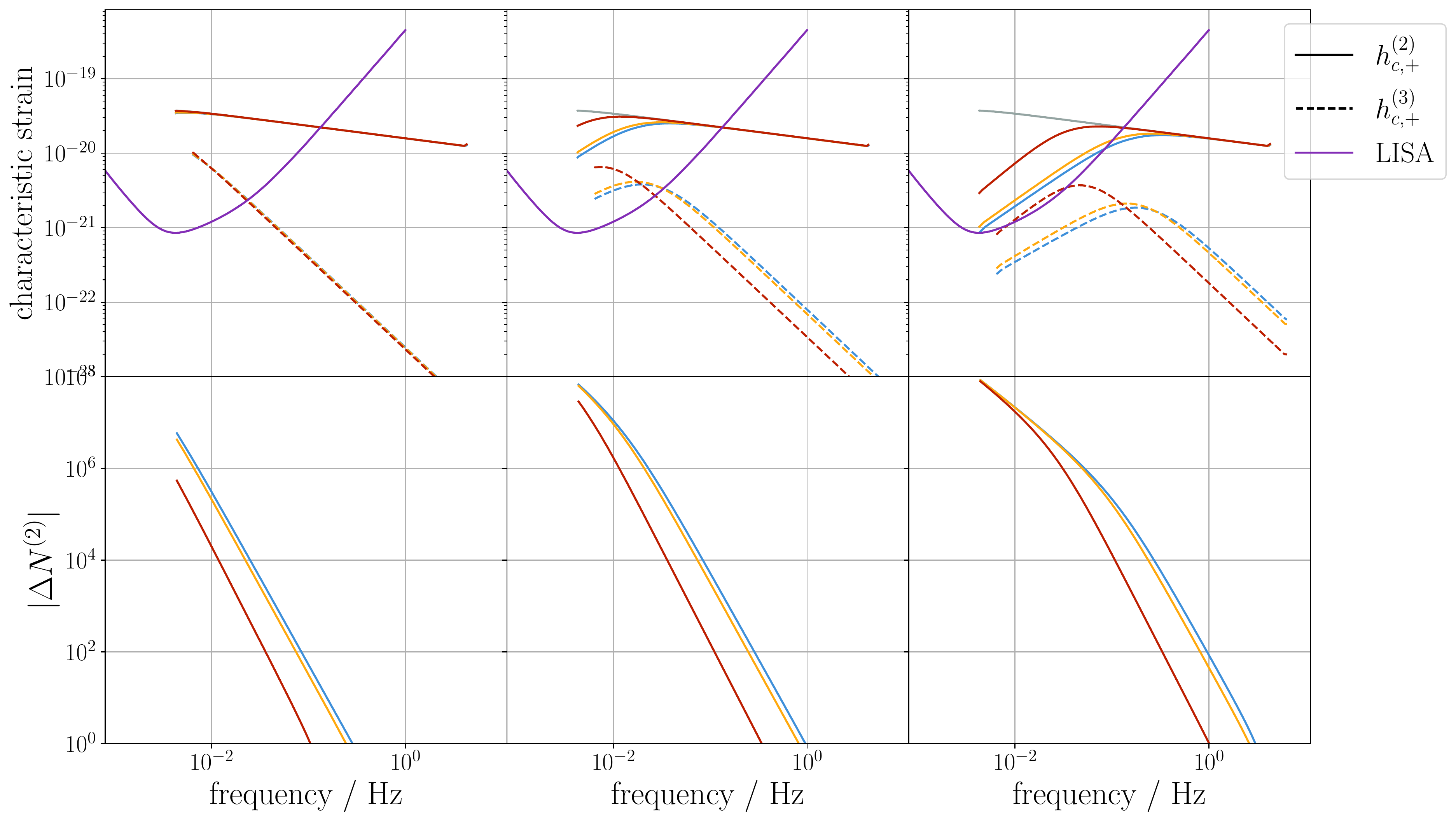}
    \caption{\textbf{Top}: The time evolution of the semimajor axis and eccentricity evolution depending on the semimajor axis. \textbf{Bottom}:  The characteristic strain of the GW signal $h^{(2)}_c$(solid), $h^{(3)}_c$(dashed) compared to the LISA sensitivity, and the dephasing $\Delta N^{(2)}$ for varying dark matter density $\rho_6=\{5.448\times 10^{13}\Msun/\pc^3, 5.448\times 10^{15}\Msun/\pc^3, 5.448\times 10^{17}\Msun/\pc^3\}$. 
    The parameters of the system are $\{D_L, m_1,m_2, \rho_6, \alpha_\sp, a_0, e_0\} = \{ 500\;\text{Mpc}, 10^3 \Msun, 1\,\Msun, 7/3, 100 \;r_\isco, 0.1\}$. 
    \label{fig:evolution_rho6}}
\end{figure*}

The next parameter we vary is the dark matter density in the form of $\rho_6$. We choose the values as $\rho_6 = \{5.448\times 10^{13}M_\odot/\pc^3, 5.448\times 10^{15}M_\odot/\pc^3, 5.448\times 10^{17}M_\odot/\pc^3 \}$ and show the results in \figref{fig:evolution_rho6}.

The time of inspiral is heavily influenced by the dark matter density. The dark matter density can speed up the inspiral from several hundred years to the order of a single year. The characteristic strain of the evolution is also strongly influenced by the dark matter density. When dynamical friction is dominant, the spectrum changes dramatically to one with a rising slope.

The evolution of the eccentricity is also dependent on the dark matter density. For $\rho_6 = 5.448\times 10^{13}M_\odot/\pc^3$ the dynamical friction effects are subdominant to the GW emission loss. Therefore, the eccentricity mostly looks like it would with just GW emission loss. For $\rho_6 = 5.448\times 10^{15}M_\odot/\pc^3$, the dynamical friction and GW emission loss are on equal footing. Early on, dynamical friction dominates, while a little later GW emission loss takes over, which can be seen in the change of shape in the characteristic strain. The eccentricity evolution is modified, and the circularization rate seems to be an average of the two effects. The difference between the different $\alpha_\sp$ models can be seen by eye. For $\rho_6 = 5.448\times 10^{17}M_\odot/\pc^3$, the behavior of the eccentricity evolution seems to flip for the different power law models. This can be explained as follows: In the next section, we will show that the circularization rate due to dynamical friction is approximately equal to $\dv{e}{t}\sim \alpha_\sp \rho_\dm$. For the lower dark matter densities either the GW emission loss dominates, or the resulting behavior is a combination of both dissipative forces. Especially for $\alpha_\sp = 7/4$, the dark matter density at $a_0\sim 100 \;r_\isco \ll 10^4 \,\isco \approx r_6$ is much lower than for the other power law indices, which is why the circularization effects are dominated by GW. For the higher dark matter density, the evolution is dominated by dynamical friction, and the circularization $\dv{e}{a}\sim \alpha_\sp$(see next section), which flips the curves.

Whether or not these dark matter densities are realized in nature remains to be seen.

Even for smaller dark matter densities, when the influence on the characteristic strain is not visible by eye, the dephasing effect is still strong with $\Delta N^{(2)}\sim 10^4-10^6$ for the different models. 

\subsubsection{Varying Central Mass}
\begin{figure*}
    \includegraphics[width=\textwidth]{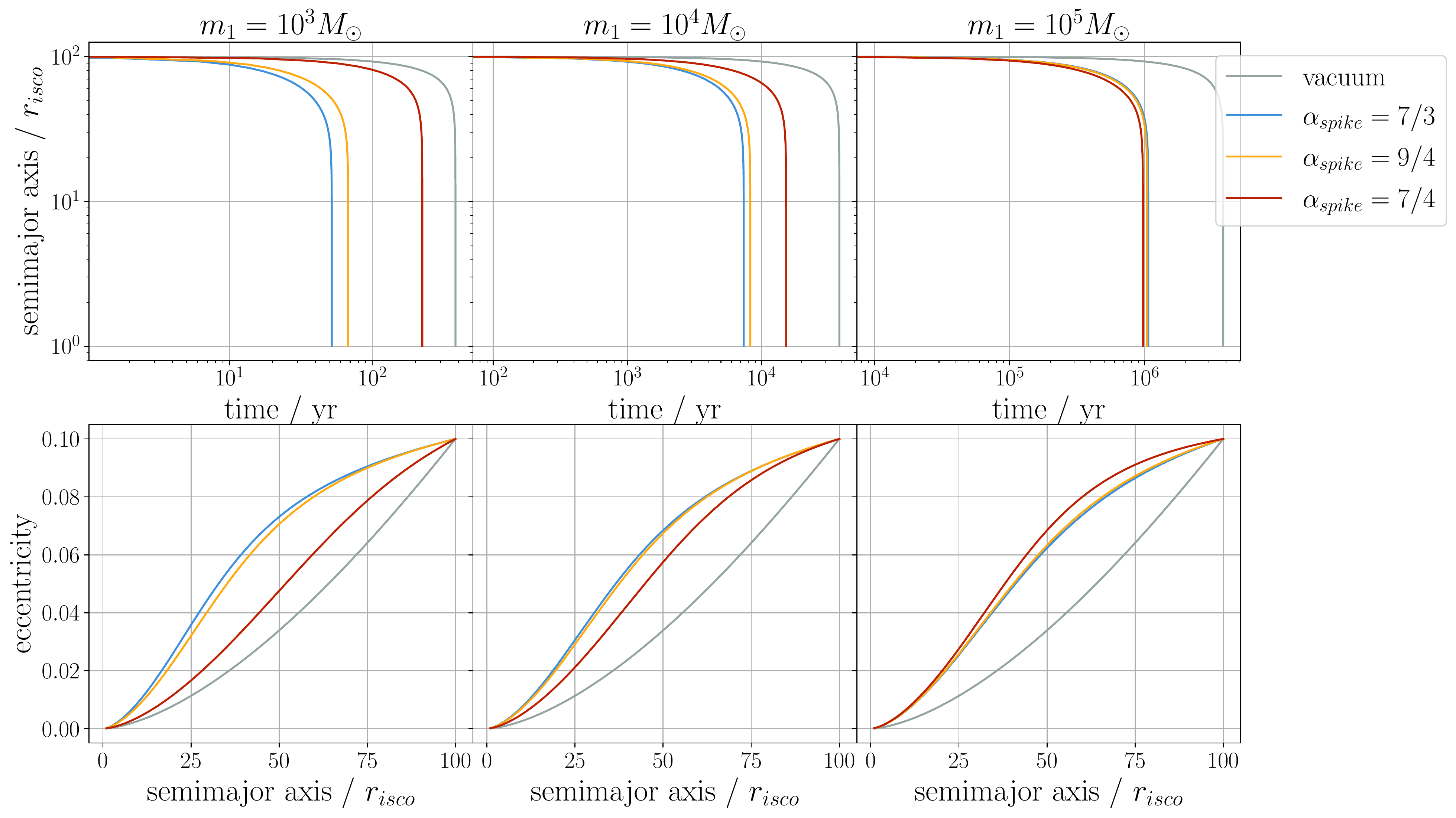}\\
    \hspace{-0.95cm}\includegraphics[width=0.955\textwidth]{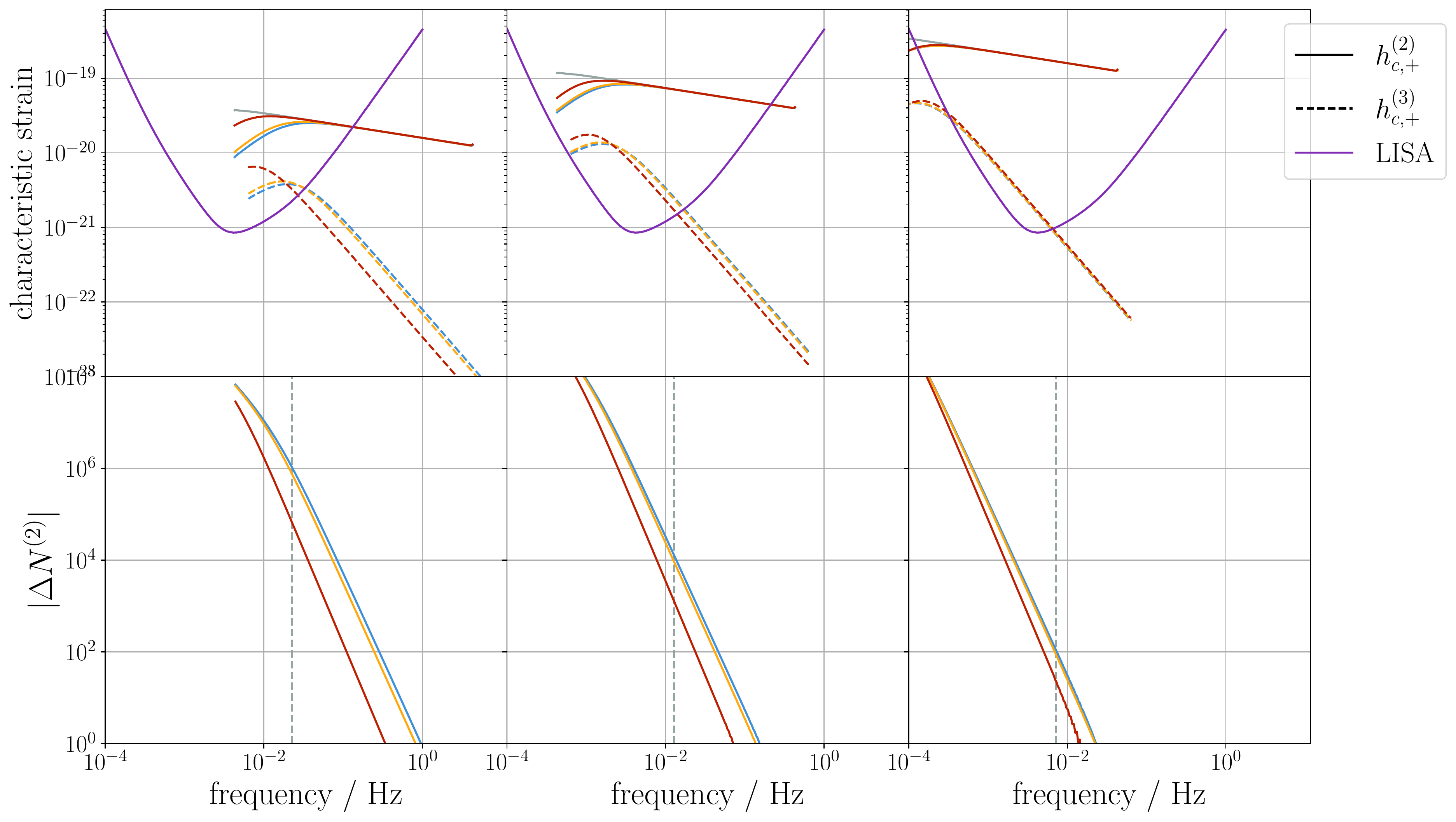}
    \caption{(\textbf{Top}: The evolution of the semimajor-axis during the inspiral. Notice the different timescales involved. \textbf{Bottom}:  The characteristic strain of the GW signal $h^{(2)}_c$(solid), $h^{(3)}_c$(dashed) compared to the LISA sensitivity, and the dephasing $\Delta N^{(2)}$ for varying central mass $m_1 = \{10^3 \Msun, 10^4 \Msun, 10^5 \Msun\}$. 
    The constant parameters of the system are $\{D_L,m_2, \rho_6, \alpha_\text{spike}, a_0, e_0\} = \{500\,\text{Mpc}, 1\,\Msun, 5.448\times 10^{15}\Msun/\pc^3, 7/3, 5\times 10^{2}\;r_\isco, 0.1\}$.
    \label{fig:evolution_m1}}
\end{figure*}

When varying the central mass, one needs to be careful about the dark matter spike density. Larger central black holes typically reside in heavier dark matter halos, which would result in a stronger spike and a higher dark matter density, see, for example, the procedure laid out in section II of \cite{Eda:2014kra}. Here, we do not consider the complications that arise from this, and instead just vary the central mass $m_1$ and see which effects it has on the model. This is shown in \figref{fig:evolution_m1}. The inspiral is considered with an initial semimajor axis of $a_0=10^2 \,r_\isco$, where $r_\isco=6\,m_1$, to allow a fair comparison between the evolution.

A similar interaction with the dark matter density can be observed here. For larger $m_1$, $r_\isco$ is larger and for $m_1 = 10^5\Msun$, $10^2 \,r_\isco \sim r_6$. Therefore, a similar flip in behavior can be seen in the evolution of the eccentricity. Since for $r > r_6$, $\alpha_\sp=7/4$ actually has the highest densities, the overall inspiral time is smaller. But this will not be observable in the dephasing signal in the last $5$ years of the system's lifetime.

As the mass of the central object $m_1$ increases, the frequency of the orbital motion and the GWs decrease, which moves the inspiral further into the LISA band. Also, the strength of the characteristic strain is increased by an order of magnitude. The inspiral takes place on much larger timescales, making it difficult to observe in its entirety. This also results in a larger total difference in the dephasing, but smaller in the last $5$ years of observation. 

Generally, a larger central mass makes the inspiral signal stronger and therefore also higher harmonics. This could make the dephasing effect and the eccentricity easier to observe. Although not observable in its entirety, it could still tease out dark matter effects. 

%% file: analysis.tex
\subsection{Conditions for circularization and eccentrification}
In this section, we further explore the eccentrification effects for a generic dissipative force and for dynamical friction specifically, with and without $\xi(v)$. 

The explanation given in the previous section -- as to why the inclusion of $\xi(v)$ circularizes the orbit -- makes sense from an orbital mechanics point of view, but not from the equations themselves. The energy loss enters into the differential equations through its average, and its shape throughout the orbit should not matter at first glance. 

Thus, to rectify this, we have to look at \eqref{eq:de_dt}, describing the evolution of the eccentricity. To this end, we first take a look at the sign of the term in the parenthesis
\begin{equation}
\label{eq:X_def}
        X := \frac{1}{E_\text{orb}} \avg{\dv{E}{t}} + 2\frac{1}{L_\text{orb}} \avg{\dv{L}{t}} 
    \begin{cases}
    < 0, \text{eccentrification} \\
    > 0, \text{circularization}
    \end{cases}
\end{equation}
First, let us focus on a single force. For a positive $X$, a single force will circularize the orbit, while for a negative $X$, it will eccentrify it. What ultimately happens to the eccentricity is then given by the relative strength of the forces. 

Plugging in Eqs. \neqref{eq:avgdEdT}, \neqref{eq:avgdLdT}, \neqref{eq:E_orbit} and \neqref{eq:L_orbit}, we can rewrite the above equation as
\begin{align}
\label{eq:X_plugged}
    X =  \frac{2(1-e^2)^{3/2} }{\mu} \int_0^{2\pi} \frac{d\phi}{2\pi} & (1+e \cos\phi)^{-2}  \nn \\
    & \times F(r,v)\left(\frac{av}{m}  - \frac{1}{v}\right).
\end{align}
To simplify calculations, we make an ansatz for the form of the force as
\begin{equation}
    F(r,v) \sim r^\alpha v^\beta,
\end{equation}
and plug in Eqs. \neqref{eq:Kepler_r} and \neqref{eq:Kepler_v}. Disregarding the prefactors, we find
\begin{align}
    X\propto  \int_0^{2\pi} \frac{d\phi}{2\pi}& (\cos\phi + e)(1+e \cos\phi)^{-(2+\alpha)} \nn \\
    & \times(1 + 2e \cos\phi + e^2)^{(\beta-1)/2} .
    \label{eq:X_approx}
\end{align}
To the first order in $e$ this integral evaluates to
\begin{align}
    X \propto& \int_0^{2\pi} \frac{d\phi}{2\pi} \left(\cos\phi + e + (-3 + \beta - \alpha) e \cos^2\phi \right ) \nn \\
     = & {}\,\, \frac{e}{2}(-1 + \beta - \alpha),
     \label{eq:X_condition}     
\end{align}
which is positive for $\alpha < \beta -1$. If this condition is fulfilled, the force will circularize the orbit. The condition also holds in third order in $e$, see Appendix \ref{sec:app_third}.

Analyzing our forces, we have
\begin{flalign*}
    F_\text{gw} \propto r^{-4}v^{-1}(11r^{-2} + v^2) &\rightarrow \text{circularization} \\ 
    F_\text{df}|_{\xi(v)\equiv1} \propto r^{-\alpha_\sp} v^{-2}  &\rightarrow \text{eccentrification for }  \alpha_\sp < 3  \\
    F_\text{df}|_{\xi(v)\propto v^3} \propto r^{-\alpha_\sp} v^{1}   &\rightarrow \text{circularization}
\end{flalign*}
The circularization of the GW emission backreaction has long been known and exploited, especially for binary systems with similar masses \cite{Maggiore:2007ulw}.

The eccentrification through dynamical friction without $\xi(v)$ has also been explored first in \cite{Yue:2019ozq}. The inclusion of $\xi(v)$ as approximately $\propto v^3$ changes the nature of the dynamical friction force, so that a power law spike must circularize the orbit. 

Thus, for $\beta = 1$, this also has the curious effect that the circularization is directly proportional to $\alpha_\sp$. Including some prefactors, we have $\dv{e}{t} \propto - X \propto - \alpha_\sp \rho_\dm(a)$. While most other effects, such as dephasing, are only affected by the local dark matter density, the circularization is sensitive to the power law index. A measurement of both the inspiral rate $\dv{a}{t}\propto \rho_\dm(a)$(in the dynamical friction dominated regime) and the circularization $\dv{e}{t}$ could in principle reveal the shape of the dark matter distribution. 

A more formal calculation gives to first order in $e$ (see Appendix \ref{sec:app_third})
\begin{equation} \label{eq:deda}
    \dv{e}{a} = \frac{e}{2a}  \, \alpha_\sp
\end{equation}
Of course, this is only valid for the idealized case where dynamical friction is the dominant force and the power law behavior of $\xi(v)$ is exact. This needs to be improved by more accurate modeling. Nevertheless, it shows that dark matter effects can be observable not just from dephasing but also from the circularization rate.

\subsection{Comparison to previous works}

A huge caveat of the approach laid out in section \ref{sec:equations} is that halo feedback is not considered. As the secondary object loses energy and angular momentum to the dark matter halo, the halo should not be considered static and evolve along with the inspiral. There can be a considerable amount of energy injected into the halo. A halo feedback model has been developed in \cite{Kavanagh:2020cfn, Coogan:2021uqv} for circular orbits. It predicts that the dark matter density would deplete locally in the region around the orbit and thus decrease dynamical friction effects. As the object inspirals, the depleted region moves inward and refills the outer region. This leaves the dark matter halo mostly intact, but prolongs the inspiral compared to the static halo case. Thus, the dephasing effect $\Delta N$ is reduced by a factor of $\sim10^2$ for a mass ratio of $q\sim10^3$. For $q\sim 10^5$ the halo feedback mechanism seems to be less relevant. 

A combination of low-eccentricity orbits and halo feedback model could see an even stronger circularization of the orbit. This is due to the fact that the orbit is locally depleted and as the object inspirals, there will generally be many more particles on the inner part compared to the outer part, strengthening the circularization effects explained in section \ref{sec:results}. Whether a circular approximation is enough to describe realistic scenarios remains to be seen. 

During the finalization of this paper, \cite{Dai:2021olt} have published their results. They include the gravitational influence of the dark matter spike as a perturbative force on osculating Keplerian orbits. This results primarily in orbital precession for large distances $p > 10^5 r_\text{isco}$ for an example case with $\{m_1, m_2, \rho_\sp, r_\sp, \alpha_\sp \} = \{10^3\Msun, 10\Msun, 226\Msun/\pc^3, 0.54\pc, 7/3 \}$. They assume that dynamical friction eccentrifies the orbit, and therefore orbital precession would be an important effect. If on the other hand the orbit is being circularized, orbital precession would be of less impact. Their observation cannot be dismissed, since on those scales, the phase space distribution function as described by \eqref{eq:rho_f} is no longer valid, because it assumes the potential to be dominated by the central black hole. We would need to model the transition phase of the potential from the central black hole to the spike dominated part to accurately assess what happens to the dynamical friction and the orbital eccentricity. 

%% file: conclusions.tex
If dark matter forms a spike around IMBHs, it will affect the inspiral of stellar mass objects around it. We have studied the dynamical friction effects in such a system. They cause a dephasing effect in the GW signal, which should be observable by LISA. We have shown that the dynamical friction losses tend to circularize the orbit, in contrast to the observations by \cite{Yue:2019ozq, Cardoso:2020iji}. This is due to the inclusion of the relative velocities of the dark matter particles. We have analyzed the mechanism behind orbital eccentrification and circularization and derived a general condition for arbitrary forces. A measurement of the circularization rate can in principle reveal the shape of the dark matter distribution. Whether the circularization effect is strong enough such that most objects in the LISA band will be circular remains to be investigated by exploring other relevant effects, such as accretion and the baryonic environment. This will be left for future work.

We have not considered the halo feedback mechanism explored in \cite{Kavanagh:2020cfn, Coogan:2021uqv}, but as the orbits are being circularized instead of eccentrified we see supporting evidence for the circular approximation they utilize.

Overall, observing the dephasing effect in an IMRI would be a unique test of the particle nature of dark matter and given the existence of dark matter spikes, should be observable with LISA.

%% file: appendix.tex
\section{Angular Momentum Loss \label{sec:app_AML}}
\eqref{eq:avgdLdT} can be derived from the relation for the specific angular momentum for Keplerian orbits
\begin{equation}
    \abs{\dv{L}{t}} = \abs{\v{r}\cross\v{F}} = \frac{\abs{\v{F}}}{\abs{\v{\dot{r}}}} \abs{\v{r}\cross \v{\dot{r}}} = \frac{F}{v}r^2 \dot{\phi},
\end{equation}
since the force vector is antiparallel to the velocity vector.\\
Together with the relation for the derivative of the true anomaly \cite{Maggiore:2007ulw}
\begin{equation}
    \dot{\phi} = \frac{\sqrt{ma(1-e^2)} }{r^2},
\end{equation}
we have
\begin{align}
    \avg{\dv{L}{t}} =& \int_0^T \frac{\mathrm{d}t}{T} \dv{L}{t} =- \int_0^T \frac{\mathrm{d}t}{T} F(r,v) \frac{r^2\dot{\phi}}{v}   \nn\\
                    =& - \sqrt{ma(1-e^2)}\int_0^T \frac{\mathrm{d}t}{T} \frac{F(r,v)}{v}.
\end{align}

\section{Condition in Third Order \label{sec:app_third}}
To third order in $e$ the integral in \eqref{eq:X_approx} evaluates to 
\begin{align}
    X \propto &-\frac{e}{16} (1 + \alpha - \beta) \\
    & \times \Big(8 + e^2 \underbrace{(11 + \alpha^2 + \alpha(7-2\beta) - 6\beta + \beta^2 )}_{Y} \Big). \nn 
\end{align}
Since $0 \leq e^2 < 1$ and $Y > -8$ for $\alpha > -10$, the condition \eqref{eq:X_condition} holds for any reasonable force. 

\section{Calculating Circularization Rate \label{sec:app_circ}}
Assuming our force to be of the form $F(r,v) = F_0 r^{\alpha} v^{\beta}$, we have 
\begin{align}
    \dv{a}{t} = & \dv{E_\text{orb}}{t} / \pdv{E_\text{orb}}{a}   \\
    = & - \frac{2}{\mu} a^{2 + \alpha - (\beta+1)/2} (1 + e^2)^{3/2 +\alpha - (\beta+1)/2} m^{(\beta-1)/2} F_0  \nn \\
    & \times \underbrace{\int_0^{2\pi} \frac{d\phi}{2\pi} (1+e \cos\phi)^{-(2+\alpha)} ( 1 + 2e \cos\phi + e^2)^{(\beta+1)/2}}_{\approx 1 + e^2/4(3 + \alpha^2 + \alpha(3-2\beta) - 2\beta + \beta^2)} \nn 
\end{align}
to second order in $e$.

Compare this to 
\begin{align}
    \dv{e}{t} =& - \frac{1-e^2}{2e}X  \\
    = & - \frac{e}{\mu}  a^{ \alpha - (\beta-1)/2} (1 - e^2)^{3/2 +\alpha - (\beta-1)/2}   \nn  \\
    & \times m^{(\beta-1)/2} F_0 (-1 + \beta - \alpha) \nn  \\
    & \times (1 + \frac{e^2}{8} (11 + \alpha^2 + \alpha(7-2\beta) - 6\beta + \beta^2) ) \nn 
\end{align}
Combining the two equations gives
\begin{align}
    \dv{e}{a} =& \frac{e(1-e^2)}{2a} \, (-1 + \beta - \alpha) \\
    &\times \frac{1 + \frac{e^2}{8} (11 + \alpha^2 + \alpha(7-2\beta) - 6\beta + \beta^2) }{1 + \frac{e^2}{4} (3 + \alpha^2 + \alpha(3-2\beta) - 2\beta + \beta^2)} \nn 
\end{align}
Neglecting the second order terms and setting $\beta=1$ and $\alpha=-\alpha_\sp$ gives \eqref{eq:deda}.